\documentclass[12pt]{article}
\usepackage{amsfonts,graphicx,amsmath,amsthm,amssymb,epsfig}
\usepackage{float}
\allowdisplaybreaks
\begin{document}

\title{\bf Extending Finch-Skea Isotropic Model to Anisotropic Domain in Modified $f(\mathcal{R},\mathcal{T})$ Gravity}
\author{Tayyab Naseer \thanks{tayyabnaseer48@yahoo.com;
tayyab.naseer@math.uol.edu.pk}~ and M. Sharif \thanks{msharif.math@pu.edu.pk} \\
Department of Mathematics and Statistics, The University of Lahore,\\
1-KM Defence Road Lahore-54000, Pakistan.}

\date{}
\maketitle

\begin{abstract}
This paper considers the Finch-Skea isotropic solution and extends
its domain to three different anisotropic interiors by using the
gravitational decoupling strategy in the context of
$f(\mathcal{R},\mathcal{T})$ gravitational theory. For this, we
consider that a static spherical spacetime is initially coupled with
the perfect matter distribution. We then introduce a Lagrangian
corresponding to a new gravitating source by keeping in mind that
this new source produces the effect of pressure anisotropy in the
parent fluid source. After calculating the field equations for the
total matter setup, we apply a transformation on the radial
component, ultimately providing two different systems of equations.
These two sets are solved independently through different
constraints that lead to some new solutions. Further, we consider an
exterior spacetime to calculate three constants engaged in the seed
Finch-Skea solution at the spherical interface. The estimated radius
and mass of a star candidate LMC X-4 are utilized to perform the
graphical analysis of the developed models. It is concluded that
only the first two resulting models are physically relevant in this
modified theory for all the considered parametric choices.
\end{abstract}

\section{Introduction}

Several discoveries have recently been made by cosmologists from
which it was shown that plenty of the astrophysical structures in
our cosmos are not randomly arranged rather they distributed in a
very organized manner. The analysis of the physical characteristics
of these interstellar bodies became a topic of significant
importance for researchers which helps to explain the phenomenon of
accelerated cosmic expansion. It was observed from different
experiments that our universe contains a big amount of a specific
force that acts opposite to gravity and results in cosmic expansion.
This force was named the dark energy because of its mysterious
nature till now. General Relativity ($\mathbb{GR}$) describes this
expanding phenomenon somehow, however, it faces some problems
related to the cosmological constant, and thus its multiple
modifications were proposed. The $f(\mathcal{R})$ theory is the
first-ever generalization among them, obtained by modifying the
Einstein-Hilbert action in which the Ricci scalar $\mathcal{R}$ and
its generic function are swapped by each other. Different cosmic
eras like inflationary and rapid expansion phases have been
discussed in this theory, and favorable outcomes were acquired
\cite{2}-\cite{4}. The self-gravitating systems have also been
discussed through several approaches in this context and found to be
physically relevant \cite{9a,9g}.

At that time, the geometry of spacetime and matter distribution were
studied independently. Initially, Bertolami et al. \cite{10}
presented that the geometry of compact celestial objects can be well
explained by taking the dependence of the above-mentioned factors on
each other. In this regard, they adopted the matter Lagrangian
depending on the scalar $\mathcal{R}$ and analyzed the effects of
such coupling on massive structures in $f(\mathcal{R})$ theory.
Several researchers have been prompted by this idea and put their
concentration to study accelerating cosmic expansion in this
modified framework. Following this, physicists thought that some new
insights regarding the evolution and expansion of our universe could
be opened once this matter-geometry coupling is generalized at the
action level. Harko et al. \cite{20} pioneered such generalization
by presenting a new theory of gravity depending on the scalar
curvature and trace of the energy-momentum tensor ($\mathbb{EMT}$)
$\mathcal{T}$, thus called it $f(\mathcal{R},\mathcal{T})$ theory.

It was observed that the considered system becomes non-conserved
under this modified theory in the light of which there appears an
extra force (influenced by different physical factors \cite{22}).
This force alters the path of moving particles from geodesic to
being in non-geodesic framework in the gravitational field. Houndjo
\cite{22a} used a minimal $f(\mathcal{R},\mathcal{T})$ model to
discuss cosmic eras like matter-dominated and late-time acceleration
phases, and concluded that this theory supports the existing
observations. Multiple minimal/non-minimal models have been proposed
in this scenario, among them $\mathcal{R}+2\xi\mathcal{T}$ gained
great attention in the literature due to its linear nature. Nashed
\cite{22b} discussed various pulsars in this linear gravity model
and solved the differential equations by making an explicit
assumption regarding the anisotropy and radial component of the
metric. They found that the estimated masses of these pulsars are in
accordance with the observed ones. Several other works on the
stellar interiors have been done in this theory from which appealing
results are obtained \cite{25ae}-\cite{25ad}.

Within the domain of astrophysics, the exploration of compact stars
stands as a captivating subject, offering valuable revelations about
the fundamental properties of matter in extreme conditions.
Discussing the isotropic models using standard approaches has been a
primary approach in comprehending these dense systems. Nonetheless,
the constraints of isotropic models become apparent when confronted
with the complexities of compact celestial objects. Various
cosmological surveys conducted in recent years have consistently
revealed deviations from isotropy within the cosmos. Notably,
investigations into inhomogeneous Supernova Ia have identified
subtle discrepancies challenging the assumption of isotropy
\cite{10aaa}-\cite{10aaa3}. Furthermore, multiple tools, including
studies of radio sources \cite{10aa}, and gamma-ray bursts
\cite{10ac}, among others, have unveiled plausible inconsistencies.
Migkas and Reiprich \cite{10a} recently discussed the cosmic
isotropy by examining the direction-dependent X-rays emanating from
massive structures. Subsequently, applying the same technique to
other galaxy clusters led to the discovery that the cosmos exhibits
anisotropic characteristics \cite{10b}. Hence, constructing
anisotropic solutions becomes imperative in gaining insights into
the cosmic origin and understanding its ultimate fate more
accurately.

Finding the solution of Einstein field equations representing
realistic self-gravitating system in the background of $\mathbb{GR}$
and modified theories becomes a crucial task for astrophysicists.
These highly non-linear equations may be solved analytically,
however, it is not possible to formulate such solutions in some
situations, leading researchers to use numerical techniques. It is
worth mentioning that the scientific community accepts only those
solutions which could be used to model physically relevant
interiors, i.e., all the required conditions are fulfilled. Finding
an appealing strategy to solve these differential equations is
itself a momentous task such that the possible resulting outcomes
would be of physical interest. In this perspective, a number of
techniques have been suggested in the literature that can be used to
study the nature of interior fluid distribution. Among them, a
particular approach prompted physicists to use it due to its novel
characteristics, referred as the gravitational decoupling. This
scheme can be employed to study a compact star possessing multiple
physical factors like density, pressure anisotropy, dissipation flux
and shear, etc., in a very convenient way.

The innovation of this approach was established on the fact that if
the field equations comprise multiple physical factors, the
gravitational decoupling divides those equations in different sets
corresponding to each matter source. These systems of equations are
easy enough to be solved independently. Gravitational decoupling
comprises two types of deformations, namely minimal and extended.
The pioneering work of Ovalle \cite{29} recently proposed the
minimal geometric deformation (MGD) strategy to formulate solutions
which are well agreed with the existing data of compact stars in the
braneworld scenario. Subsequently, Ovalle and Linares \cite{30}
considered an isotropic self-gravitating spherical fluid interior
and formulated the corresponding exact solution, showing consistency
with Tolman-IV metric. Casadio et al. \cite{31} extended this scheme
to obtain the Schwarzschild spacetime in the braneworld.

An isotropic sphere has been taken into account by Ovalle and his
collaborators \cite{33} whose domain was extended to anisotropic
using MGD. Sharif and Sadiq \cite{34} extended these solutions to
the Einstein-Maxwell framework to obtain two anisotropic
counterparts of isotropic Krori-Barua metric and explored the
effects of decoupling and charge on them. Gabbanelli et al.
\cite{36} generalized the isotropic Durgapal-Fuloria metric to
anisotropic domain via MGD and concluded that the resulting models
represent physically realistic compact star. A similar investigation
has also been done for the isotropic Heintzmann and Tolman VII
ansatz from which acceptable outcomes were obtained for particular
parametric choices \cite{36a,37a}. We have used the Krori-Barua
ansatz to develop anisotropic uncharged/charged self-gravitating
models and checked the impact of non-minimal fluid-geometry
interaction as well as decoupling parameter on them
\cite{37f}-\cite{37h}.

Scientific literature has introduced numerous methodologies to
tackle the challenges appear when solving Einstein/modified field
equations. These approaches include employing specific equations of
state, known metric components, or imposing conditions of vanishing
complexity. In the domain of radial and time metric components, the
Finch-Skea ansatz \cite{38ca} has attracted significant attention
from researchers as a valuable tool for exploring anisotropic
compact structures. Notably, Banerjee et al. \cite{38m} introduced a
family of interior solutions analogous to the BTZ exterior by
employing this metric. Their work yielded physically viable results,
even in lower dimensions, demonstrating the versatility and
applicability of the considered ansatz. Researchers have introduced
three distinct analytic solutions to the charged field equations,
verifying their adherence to stability criteria \cite{38n}. The
subsequent physical analysis suggests that this analog can
effectively model compact stars in the presence of Maxwell field
with realistic physical properties. Various approaches, such as
gravitational decoupling and considering complexity factors, have
been employed to discuss self-gravitating charged or uncharged
systems \cite{38o}-\cite{38q}.

In this article, we formulate multiple extensions of an isotropic
Finch-Skea metric to the anisotropic domain by the help of a
systematic technique in the framework of modified gravity. The
arrangement of this paper can be understood in the following lines.
Section \textbf{2} defines some fundamentals of
$f(\mathcal{R},\mathcal{T})$ theory and determines the independent
components of the field equations representing the total fluid
source (perfect and newly added). We then introduce the MGD scheme
in section \textbf{3} whose implementation on the field equations
divides them into two distinct sets. Section \textbf{4} is devoted
to the computation of an unknown triplet in the Finch-Skea solution
through junction conditions. A physically relevant model must
satisfy some requirements that are presented in section \textbf{5}.
Finally, we formulate three new anisotropic solutions alongside
their graphical interpretation in section \textbf{6} and summarize
our results in section \textbf{7}.

\section{$f(\mathcal{R},\mathcal{T})$ Gravity}

The action for $f(\mathcal{R},\mathcal{T})$ gravity can be defined
in the presence of an extra matter field as \cite{20}
\begin{equation}\label{g1}
S=\int \sqrt{-g}\left[\frac{f(\mathcal{R},\mathcal{T})}{16\pi}
+\L_{m}+\zeta\L_{\mathrm{E}}\right]d^{4}x,
\end{equation}
where $\L_{m}$ is the Lagrangian density of the fluid distribution
and $\L_{\mathrm{E}}$ is the additional gravitationally coupled
fluid source. Here, $\zeta$ represents the decoupling parameter that
examines how the new source affects the original matter field. The
modified field equations can be obtained by varying the action
\eqref{g1} with respect to the metric tensor. These equations are
given by
\begin{equation}\label{g2}
\mathbb{G}_{\sigma\omega}=8\pi
\mathcal{T}_{\sigma\omega}^{\mathrm{(tot)}},
\end{equation}
where $\mathbb{G}_{\sigma\omega}$ is called the Einstein tensor
representing geometry of the structure and
$\mathcal{T}_{\sigma\omega}^{\mathrm{(tot)}}$ denotes the fluid
distribution comprised by the interior of that spacetime. We
classify the later term into three different segments as
\begin{equation}\label{g3}
\mathcal{T}_{\sigma\omega}^{\mathrm{(tot)}}=\frac{1}{f_{\mathcal{R}}}\big(\mathcal{T}_{\sigma\omega}+\zeta
\mathrm{E}_{\sigma\omega}\big)+\mathcal{T}_{\sigma\omega}^{(C)}.
\end{equation}
Here, the first two terms on the right side are the usual and
additional fluid $\mathbb{EMT}$s, respectively. Also, the term
$\mathcal{T}_{\sigma\omega}^{(C)}$ shows the corrections of the
modified theory, defined as
\begin{eqnarray}
\nonumber \mathcal{T}_{\sigma\omega}^{(C)}&=&\frac{1}{8\pi
f_{\mathcal{R}}}\bigg[f_{\mathcal{T}}\mathcal{T}_{\sigma\omega}+\bigg\{\frac{\mathcal{R}}{2}\bigg(\frac{f}{\mathcal{R}}
-f_{\mathcal{R}}\bigg)-\L_{m}f_{\mathcal{T}}\bigg\}g_{\sigma\omega}\\\label{g4}
&-&(g_{\sigma\omega}\Box-\nabla_{\sigma}\nabla_{\omega})f_{\mathcal{R}}+2f_{\mathcal{T}}g^{\zeta\beta}\frac{\partial^2
\L_{m}}{\partial g^{\sigma\omega}\partial g^{\zeta\beta}}\bigg],
\end{eqnarray}
where $f_{\mathcal{T}}=\frac{\partial
f(\mathcal{R},\mathcal{T})}{\partial \mathcal{T}}$ and
$f_{\mathcal{R}}=\frac{\partial f(\mathcal{R},\mathcal{T})}{\partial
\mathcal{R}}$. Moreover,
$\Box\equiv\frac{1}{\sqrt{-g}}\partial_\sigma\big(g^{\sigma\omega}\sqrt{-g}\partial_{\omega}\big)$
is the D'Alembert operator and $\nabla_\sigma$ symbolizes the
covariant divergence.

It is important to mention here that there are two fluid sources in
the interior of a compact geometry, one is initial (which we can
call as ``seed'') source and we consider it to be the isotropic
perfect fluid at some initial moment. On the other hand, the second
matter source is known as extra or additional fluid that produces
anisotropy in the initial (seed) source. The $\mathbb{EMT}$
corresponding to the isotropic fluid is defined by
\begin{equation}\label{g5}
\mathcal{T}_{\sigma\omega}=(\mu+P)\mathcal{K}_{\sigma}\mathcal{K}_{\omega}+P
g_{\sigma\omega},
\end{equation}
where $P,~\mu$ and $\mathcal{K}_{\sigma}$ indicate the isotropic
pressure, energy density and four-velocity, respectively. Also, the
trace of Eq.\eqref{g2} is obtained as
\begin{align}\nonumber
&\mathcal{R}f_\mathcal{R}+3\nabla^{\sigma}\nabla_{\sigma}f_\mathcal{R}2f-\mathcal{T}(f_\mathcal{T}+1)-\zeta\mathrm{E}
+4f_\mathcal{T}\L_m-2f_\mathcal{T}g^{\zeta\beta}g^{\sigma\omega}\frac{\partial^2\L_m}{\partial
g^{\zeta\beta}\partial g^{\sigma\omega}}=0.
\end{align}
When we assume a vacuum case (i.e., $\mathcal{T}=0$), the
fluid-geometry interaction vanishes and the results of this theory
reduce to $f(\mathcal{R})$ framework. Moreover, we observe some
extra terms in the divergence of $\mathbb{EMT}$ due to such
interaction, leading to the non-conservation phenomenon. This
non-conservation provides an evident for the existence of an
additional force that can mathematically be expressed by
\begin{align}\nonumber
\nabla^\sigma\mathcal{T}_{\sigma\omega}&=\frac{f_\mathcal{T}}{8\pi-f_\mathcal{T}}\bigg[(\mathcal{T}_{\sigma\omega}
+\Upsilon_{\sigma\omega})\nabla^\sigma\ln{f_\mathcal{T}}
-\frac{1}{2}g_{\alpha\beta}\nabla_\omega\mathcal{T}^{\alpha\beta}\\\label{g11}
&+\nabla^\sigma\bigg(g_{\sigma\omega}\L_m-2\mathcal{T}_{\sigma\omega}-2g^{\zeta\beta}\frac{\partial^2
\L_{m}}{\partial g^{\sigma\omega}\partial
g^{\zeta\beta}}\bigg)-\frac{8\pi\zeta}{f_\mathcal{T}}\nabla^\sigma\mathrm{E}_{\sigma\omega}\bigg].
\end{align}
The self-gravitating geometry is usually divided into two regions,
namely interior and exterior, and both of them are distinguished
from each other at some boundary called the hypersurface ($\Sigma$).
The following metric describes the interior spacetime given as
\begin{equation}\label{g6}
ds^2=-e^{a_1} dt^2+e^{a_2} dr^2+r^2\big(d\theta^2+\sin^2\theta
d\phi^2\big),
\end{equation}
where $a_1=a_1(r)$ and $a_2=a_2(r)$. The four-velocity defined in
Eq.\eqref{g5} now becomes in relation with the metric \eqref{g6} as
\begin{equation}\label{g7}
\mathcal{K}_\sigma=-\delta^0_\sigma
e^{\frac{a_1}{2}}=(-e^{\frac{a_1}{2}},0,0,0).
\end{equation}
To understand the physical behavior of solutions to the field
equations is indispensable requirement so that it can be explored
whether they represent physically realistic compact models or not.
This cannot be done without taking a standard model of the
considered gravity theory into account. There are several models
proposed in the literature, however, we adopt the simplest one among
them so that the geometric deformations can effectively be employed
on the corresponding field equations. The model under consideration
is given as follows
\begin{equation}\label{g7a}
f(\mathcal{R},\mathcal{T})=f_1(\mathcal{R})+
f_2(\mathcal{T})=\mathcal{R}+2\xi\mathcal{T},
\end{equation}
where $\xi$ symbolizes an arbitrary constant and
$\mathcal{T}=-\mu+3P$. Multiple potentials showing inflation have
been adopted in this context to derive potential slow-roll
parameters. It was then concluded that the resulting outcomes are
consistent with the observed data only for a particular range of
$\xi\in(-0.37,1.483)$  \cite{38}. The Tolman-Kuchowicz interior
possessing anisotropic fluid distribution has also been successfully
developed by using the above linear model \cite{39}. We have
utilized the same model to formulate several acceptable compact
interiors through the complexity-free constraint and a liner
equation of state \cite{40,40b}.

Combining $f(\mathcal{R},\mathcal{T})$ model \eqref{g7a} with the
general field equations \eqref{g6}, the non-vanishing components are
calculated as
\begin{align}\label{g8}
&e^{-a_2}\left(\frac{a_2'}{r}-\frac{1}{r^2}\right)
+\frac{1}{r^2}=8\pi\left(\mu-\zeta\mathrm{E}_{0}^{0}\right)+\xi\left(3\mu-P\right),\\\label{g9}
&e^{-a_2}\left(\frac{1}{r^2}+\frac{a_1'}{r}\right)
-\frac{1}{r^2}=8\pi\left(P+\zeta\mathrm{E}_{1}^{1}\right)-\xi\left(\mu-3P\right),
\\\label{g10}
&\frac{e^{-a_2}}{4}\left[a_1'^2-a_2'a_1'+2a_1''-\frac{2a_2'}{r}+\frac{2a_1'}{r}\right]
=8\pi\left(P+\zeta\mathrm{E}_{2}^{2}\right)-\xi\left(\mu-3P\right),
\end{align}
where the matter terms multiplied by $\xi$ are the corrections of
modified theory and $'=\frac{\partial}{\partial r}$. Since the
spherical system now possesses two fluid sources, the
Tolman-Opphenheimer-Volkoff equation \eqref{g11} can be generalized
as
\begin{align}\nonumber
&\frac{dP}{dr}+\frac{a_1'}{2}\left(\mu+P\right)+\frac{\zeta a_1'}{2}
\left(\mathrm{E}_{1}^{1}-\mathrm{E}_{0}^{0}\right)+\zeta\frac{d\mathrm{E}_{1}^{1}}{dr}\\\label{g12}
&+\frac{2\zeta}{r}\left(\mathrm{E}_{1}^{1}-\mathrm{E}_{2}^{2}\right)=-\frac{\xi}{4\pi-\xi}\big(\mu'-P'\big).
\end{align}
The non-zero right hand side of the above equation makes this theory
non-conserved, however, $\xi=0$ would lead to the conservation
equation. This equation is, in fact, a sum of different physical
forces which help to keep a self-gravitating system in an
equilibrium state. We observe that the number of unknowns in
Eqs.\eqref{g8}-\eqref{g10} are now increased due to the involvement
of components of an extra source, i.e.,
$(a_1,a_2,\mu,P,\mathrm{E}_{0}^{0},\mathrm{E}_{1}^{1},\mathrm{E}_{2}^{2})$.
Therefore, a systematic approach \cite{33} must be required to work
out this under-determined system.

\section{Minimal Gravitational Decoupling}

Gravitational decoupling is the most advantageous technique that
divides the field equations into different sets by deforming the
metric components from one reference frame to another, making it
easy to solve these distinct systems independently. Since we need a
new reference frame in which the field equations
\eqref{g8}-\eqref{g10} shall be transformed, we assume another line
element given as follows
\begin{equation}\label{g15}
ds^2=-e^{a_3(r)}dt^2+\frac{1}{a_4(r)}dr^2+r^2\big(d\theta^2+\sin^2\theta
d\phi^2\big).
\end{equation}
The transformation equations are defined as
\begin{equation}\label{g16}
a_3\rightarrow a_1=a_3+\zeta\mathrm{t}_1^*, \quad a_4\rightarrow
e^{-a_2}=a_4+\zeta\mathrm{t}_2^*,
\end{equation}
where $\mathrm{t}_1^*$ and $\mathrm{t}_2^*$ correspond to time and
radial components, respectively. We consider MGD strategy in the
current setup that allows only the transformation of $g_{rr}$
potential whereas $g_{tt}$ component remains the same, i.e.,
$\mathrm{t}_1^*\rightarrow 0,~\mathrm{t}_2^*\rightarrow
\mathrm{T}^*$. Accordingly, Eq.\eqref{g16} turns into
\begin{equation}\label{g17}
a_3\rightarrow a_1=a_3, \quad a_4\rightarrow
e^{-a_2}=a_4+\zeta\mathrm{T}^*,
\end{equation}
where $\mathrm{T}^*=\mathrm{T}^*(r)$. While using these
transformations on the field equations, we must keep in mind that
the spherical symmetry is not disturbed by them. We now apply a
linear mapping \eqref{g17} on Eqs.\eqref{g8}-\eqref{g10}, the first
resulting set characterizing the perfect fluid source is obtained as
\begin{align}\label{g18}
&e^{-a_2}\left(\frac{a_2'}{r}-\frac{1}{r^2}\right)
+\frac{1}{r^2}=8\pi\mu+\xi\left(3\mu-P\right),\\\label{g19}
&e^{-a_2}\left(\frac{1}{r^2}+\frac{a_1'}{r}\right)
-\frac{1}{r^2}=8\pi P-\xi\left(\mu-3P\right),\\\label{g20}
&\frac{e^{-a_2}}{4}\left[a_1'^2-a_2'a_1'+2a_1''-\frac{2a_2'}{r}+\frac{2a_1'}{r}\right]
=8\pi P-\xi\left(\mu-3P\right).
\end{align}
Since the above system represents isotropic matter source, the first
two field equations can completely define the corresponding
interior. Hence, Eqs.\eqref{g18} and \eqref{g19} produce the
explicit form of the energy density and pressure as
\begin{align}\label{g18a}
\mu&=\frac{e^{-a_2}}{8r^2\big(\xi^2+6\pi\xi+8\pi^2\big)}\big[\xi r
a_1 '+(3 \xi +8 \pi ) r a_2 '+2 (\xi +4 \pi ) \big(e^{a_2
}-1\big)\big],\\\label{g19a}
P&=\frac{e^{-a_2}}{8r^2\big(\xi^2+6\pi\xi+8\pi^2\big)}\big[(3 \xi +8
\pi ) r a_1 '+\xi  r a_2 '-2 (\xi +4 \pi ) \big(e^{a_2
}-1\big)\big].
\end{align}

The second set, on the other hand, is characterizing the newly added
fluid source and is given as follows
\begin{align}\label{g21}
&8\pi\mathrm{E}_{0}^{0}=\frac{\mathrm{T}^{*'}}{r}+\frac{\mathrm{T}^*}{r^2},\\\label{g22}
&8\pi\mathrm{E}_{1}^{1}=\mathrm{T}^*\left(\frac{a_1'}{r}+\frac{1}{r^2}\right),\\\label{g23}
&8\pi\mathrm{E}_{2}^{2}=\frac{\mathrm{T}^*}{4}\left(2a_1''+a_1'^2+\frac{2a_1'}{r}\right)
+\mathrm{T}^{*'}\left(\frac{a_1'}{4}+\frac{1}{2r}\right).
\end{align}
The interesting fact of the MGD scheme is that switching energy and
momentum from the original source to the seed one and vice-versa is
not allowed, resulting in the conservation of both sources
individually. It becomes easy for us now to solve both sets of field
equations independently and then add their solutions in a particular
way to get the solution of total fluid setup. In the system
\eqref{g18a} and \eqref{g19a}, four unknowns ($\mu,P,a_1,a_2$) are
appeared which can be tackled by making any two factors known
through some constraints. On the other hand, the system
\eqref{g21}-\eqref{g23} also engages four unknowns
($\mathrm{T}^*,\mathrm{E}_{0}^{0},\mathrm{E}_{1}^{1},\mathrm{E}_{2}^{2}$),
thus only one constraint is needed to obtain a unique solution.
Since we consider an extra fluid setup as a source of generating
anisotropy in the original perfect distribution, the matter
determinants must be identified as follows
\begin{equation}\label{g13}
\tilde{\mu}=\mu-\zeta\mathrm{E}_{0}^{0},\quad
\tilde{P}_{r}=P+\zeta\mathrm{E}_{1}^{1}, \quad
\tilde{P}_{\bot}=P+\zeta\mathrm{E}_{2}^{2},
\end{equation}
along with the anisotropic factor defined by
\begin{equation}\label{g14}
\tilde{\Pi}=\tilde{P}_{\bot}-\tilde{P}_{r}=\zeta(\mathrm{E}_{2}^{2}-\mathrm{E}_{1}^{1}),
\end{equation}
verifying that extracting the impact of a new fluid source (i.e.,
$\zeta=0$) again leads to the parent (isotropic) source.

\section{Finch-Skea Metric and Boundary Conditions}

We obtain a unique solution of Eqs.\eqref{g18a} and \eqref{g19a} in
this section. It is already discussed that if we adopt two
constraints or a particular metric ansatz, then the four unknowns
can easily be evaluated. Since we are dealing with perfect
distribution, we adopt the Finch-Skea isotropic ansatz \cite{42a}
given by the metric
\begin{align}\label{g33}
ds^2=-\frac{1}{4}\big(2C_1+C_2\sqrt{C_3}r^2\big)^2dt^2+\big(C_3r^2+1\big)dr^2+r^2\big(d\theta^2+\sin^2\theta
d\phi^2\big),
\end{align}
where a triplet ($C_1,C_2,C_3$) is needed to determine so that the
graphical interpretation of the proposed models in the following
shall be entertained. Their dimensions are null, $\frac{1}{\ell}$
and $\frac{1}{\ell^2}$, respectively. The metric \eqref{g33}
produces the energy density \eqref{g18a} and isotropic pressure
\eqref{g19a} as
\begin{align}\nonumber
\mu&=\frac{1}{4r^2 (\xi +2 \pi ) (\xi +4 \pi ) \left(C_3
r^2+1\right)}\\\label{g35} &\times\bigg\{5 \xi +(\xi +4 \pi ) C_3
r^2-\frac{4 \xi  C_1}{C_2 \sqrt{C_3} r^2+2 C_1}-\frac{3 \xi +8 \pi
}{C_3 r^2+1}+8 \pi\bigg\},\\\nonumber P&=\frac{-1}{4r^2 (\xi +2 \pi
) (\xi +4 \pi ) \left(C_3 r^2+1\right)}\\\label{g36}
&\times\bigg\{(\xi +4 \pi ) C_3 r^2+\xi  \left(\frac{1}{C_3
r^2+1}-7\right)+\frac{4 (3 \xi +8 \pi ) C_1}{C_2 \sqrt{C_3} r^2+2
C_1}-16 \pi\bigg\}.
\end{align}

In order to evaluate three constants, we need an exterior metric to
match with the interior spherical spacetime at the interface
($\Sigma:~r=R$). Moreover, the junction conditions help to
understand the complete structure of a geometrical system. The
interior geometry is given in Eq.\eqref{g33}, however, the most
suitable exterior region must be vacuum, representing by the
Schwarzschild solution. It is important to note the distinctions in
junction conditions between $\mathbb{GR}$ and $f(\mathcal{R})$
theory due to the inclusion of higher-order geometric terms
\cite{a2,a3}. However, in the vacuum scenario, the term
$\mathcal{T}$ in the model \eqref{g7a} has null contribution.
Consequently, the exterior line element can be adopted similarly to
that of $\mathbb{GR}$. We express its metric in the following
\begin{equation}\label{g25}
ds^2=-\bigg(1-\frac{2M}{r}\bigg)dt^2+\bigg(1-\frac{2M}{r}\bigg)^{-1}dr^2+
r^2\big(d\theta^2+\sin^2\theta d\phi^2\big),
\end{equation}
where $M$ being the total exterior mass. We consider the metric
potentials of both geometries and their derivatives to be continuous
across the boundary that can mathematically be expressed as
\begin{equation}\label{g34}
g_{tt}^-~{_=^\Sigma}~g_{tt}^+, \quad g_{rr}^-~{_=^\Sigma}~g_{rr}^+,
\quad g_{tt,r}^-~{_=^\Sigma}~g_{tt,r}^+,
\end{equation}
where plus and minus signs correspond to outer and inner regions,
respectively. The above three constraints provide the following
equations as
\begin{eqnarray}\label{g36ab}
\frac{R-2M}{R}&=&\frac{1}{4}\big(2C_1+C_2\sqrt{C_3}r^2\big)^2,\\\label{g36ac}
\frac{R}{R-2M}&=&C_3r^2+1,\\\label{g36aca}
\frac{2M}{R^2}&=&C_2\big(2C_1\sqrt{C_3}r+C_2C_3r^3\big).
\end{eqnarray}

We simultaneously solve Eqs.\eqref{g36ab}-\eqref{g36aca} to get the
following constants as
\begin{eqnarray}\label{g37}
C_1&=&\frac{2R-5M}{2R\sqrt{R-2M}},\\\label{g38}
C_2&=&\frac{1}{R^{\frac{3}{2}}}\sqrt{\frac{M}{2}},\\\label{g38a}
C_3&=&\frac{2M}{R^2\big(R-2M\big)}.
\end{eqnarray}
Another way to find these constants is to assume the first two
constraints given in Eq.\eqref{g34} along with the disappearing
pressure at the spherical boundary (i.e., $P~{_=^\Sigma}~0$). These
three equations would generate different values of the triplet which
can also be used to perform the graphical analysis.

\section{Physical Constraints on Compact Models}

The literature is full of exact/numerical solutions to the field
equations in GR or any other modified gravity. So the question can
arise in one's mind whether all these solutions are of interest for
the scientific community or not, however, the answer is no.
Researchers \cite{ab,ac} provided multiple constraints that must be
implemented on the compact models to check whether they are
physically realistic or not. Thus, the only solutions fulfilling the
following constraints are significant.
\begin{itemize}
\item Existing the singularity inside a compact structure is the first primary factor
that demands our attention. Therefore, our investigation requires to
confirm the non-singular nature of both metric components of
Finch-Skea ansatz. Ensuring their increasing nature outwards is also
a crucial requirement to obtain a viable self-gravitating structure.
In the following, we check the behavior of time/radial components of
the metric \eqref{g33} at the center.
\begin{align}\nonumber
e^{a_1(r)}|_{r=0}=C_1^2,\quad e^{a_2(r)}|_{r=0}=1.
\end{align}
The first derivative of these potentials is
\begin{align}\nonumber
(e^{a_1(r)})'=2 C_2 \sqrt{C_3} r \left(\frac{1}{2} C_2 \sqrt{C_3}
r^2+C_1\right), \quad (e^{a_2(r)})'=2C_3r,
\end{align}
and we find them zero at $r=0$, identifying the increasing profile
of both components, hence, the Finch-Skea ansatz is valid to be
used.

\item It is imperative to study the nature of fluid-related
parameters such as pressure and energy density. These components
must exhibit finite and positive values throughout the entire
domain. Further, they must be maximum (or minimum) at the center (or
the outer boundary $\Sigma:~r=R$). Likewise, we can say that their
decreasing trend towards the boundary can be guaranteed if the first
derivatives of these parameters disappear at $r=0$ while showing a
negative gradient outwards.

\item The spherical system possesses a mass that can be evaluated in terms of both geometry and fluid distribution.
Here, we calculate the mass of the considered setup using the later
definition. Its mathematical description is
\begin{equation}\label{g39}
m(r)=\frac{1}{2}\int_{0}^{R}w^2\mu dw.
\end{equation}
One more crucial concept in the field of astrophysics, particularly
in the study of compact stars, is the compactness factor that
measures the tight concentration of mass within the given region. It
is also expressed as mass to the size (radius) ratio of a compact
body, i.e., $\nu(r)=\frac{m(r)}{r}$. A structure having a high
compactness value describes that a large amount of matter is packed
into a very small volume. Such objects produce robust gravitational
effects. For the case of a spherical body, the compactness has been
found to be less than $\frac{4}{9}$ \cite{42a}. A self-gravitating
object's surface emits light/electromagnetic radiations due to the
strong gravitational effects of other near by bodies. A phenomenon
that observes the change in the wavelength of those radiations is
called surface redshift and we can define it in terms of the
compactness as
\begin{equation}\label{g41}
\mathrm{z}(r)=\frac{1-\sqrt{1-2\nu(r)}}{\sqrt{1-2\nu(r)}}.
\end{equation}
It is important to highlight that the surface redshift attains its
maximum value of $5.211$ in the context of anisotropic fluid
distribution \cite{42b} whereas it becomes less for an isotropic
configuration.

\item The viability of the interior compact fluid configuration
can be checked the energy conditions. They are, in fact, linear
combinations of pressure and the density reigning the
isotropic/anisotroipc setup. Ensuring their positive profile
throughout the domain guarantees the viability of the developed
model and thus the existence of usual matter is verified. If any of
them is not satisfied, an exotic fluid exists. In relation with
anisotropic interior, these conditions are
\begin{eqnarray}\nonumber
&&\mu \geq 0, \quad \mu+P_{\bot} \geq 0, \quad \mu+P_{r} \geq
0,\\\label{g50} &&\mu-P_{\bot} \geq 0, \quad \mu-P_{r} \geq 0, \quad
\mu+2P_{\bot}+P_{r} \geq 0.
\end{eqnarray}
The positive behavior of the physical determinants demands only the
satisfaction of the dominant bounds (i.e., $\mu-P_{r} \geq 0$ and
$\mu-P_{\bot} \geq 0$ implying $\mu \geq P_{\bot}$ and $\mu \geq
P_{r}$, respectively) as all other conditions would be trivially
fulfilled.

\item A celestial body may possess several components due to which
there occur fluctuations in its interior, departing that body from
the state of hydrostatic equilibrium. This leads to the instability
of the considered fluid distribution and thus the long-term
existence of celestial objects may not take place. In this regard,
multiple approaches based on the sound speed and perturbation
methods have been suggested to analyze the structural stability. It
has been found that the speed of light must be greater than the
sound speed only if the body is stable. Mathematically,
$$0 < v_{sr}^{2}=\frac{dP_{r}}{d\mu},~ v_{s\bot}^{2}=\frac{dP_{\bot}}{d\mu} < 1,$$
where $v_{sr}^{2}$ and $v_{s\bot}^{2}$ are the radial and tangential
speeds of sound, respectively \cite{42bb}. In a similar fashion,
Herrera \cite{42ba} proposed that the cracking can be occurred if
the total force in the radial direction changes its sign. Avoiding
such cracking is necessary to obtain a physically stable model. It
is found that the factor $|v_{s\bot}^{2}-v_{sr}^{2}|$ must lie
between $-1$ and $1$ in order to avoid the cracking.
\end{itemize}

\section{Some New Anisotropic Solutions}

In the last couple of years, many researchers are actively involved
in finding the solutions of the field equations that characterize
the interior matter distribution in different theories of gravity.
For this purpose, several constraints/schemes have been suggested,
resulting in physically acceptable compact celestial structures.
Since we are left to solve the system \eqref{g21}-\eqref{g23}
containing four unknowns, only one constraint is required at a time
to find the corresponding unique solution. In this paper, we choose
the following constraints
\begin{itemize}
\item Density-like constraint,
\item Pressure-like constraint,
\item A linear equation of state.
\end{itemize}

\subsection{Model I}

According to the density-like constraint, the energy density of both
the matter sources can be put equal to each other \cite{42baba}.
This provides several anisotropic physically relevant compact
interiors in $\mathbb{GR}$ and modified theories. We define it as
follows
\begin{equation}\label{g51}
\mu=\mathrm{E}_{0}^{0}.
\end{equation}
Substituting Eqs.\eqref{g18a} and \eqref{g22} in the above equation,
we get the linear-order differential equation given by
\begin{align}\nonumber
&\frac{1}{8\pi}\bigg\{\frac{\mathrm{T}^{*'}(r)}{r}+\frac{\mathrm{T}^*(r)}{r^2}\bigg\}
-\frac{e^{-a_2}}{8r^2\big(\xi^2+6\pi\xi+8\pi^2\big)}\\\label{g52}
&\times\big[\xi r a_1 '+(3 \xi +8 \pi ) r a_2 '+2 (\xi +4 \pi )
\big(e^{a_2 }-1\big)\big]=0,
\end{align}
that becomes after combining with the metric \eqref{g33} as
\begin{align}\nonumber
&\frac{1}{8\pi}\bigg\{\frac{\mathrm{T}^{*'}(r)}{r}+\frac{\mathrm{T}^*(r)}{r^2}\bigg\}-\frac{1}{4r^2
(\xi +2 \pi ) (\xi +4 \pi ) \left(C_3 r^2+1\right)}\\\label{g53}
&\times\bigg\{5 \xi +(\xi +4 \pi ) C_3 r^2-\frac{4 \xi  C_1}{C_2
\sqrt{C_3} r^2+2 C_1}-\frac{3 \xi +8 \pi }{C_3 r^2+1}+8
\pi\bigg\}=0.
\end{align}
This equation involves only one unknown as the deformation function,
thus we can easily solve it. It is not possible sometimes to
calculate the exact solutions of the differential equations due to
the involvement of complex functions. However, in this case, we
obtain an analytic solution successfully. The deformation function
is obtained as
\begin{align}\nonumber
\mathrm{T}^*(r)&=\frac{\pi }{rC_3^{5/4}\big(\xi ^2+6 \pi  \xi +8 \pi
^2\big) \big(2 C_1 \sqrt{C_3}-C_2\big)}\bigg[2rC_3^{5/4} \big(2 C_1
\sqrt{C_3}-C_2\big)\\\nonumber &\times (\xi +4 \pi )+4
\sqrt{2C_1C_2} \xi C_3 \tan ^{-1}\big\{\sqrt{\frac{C_2}{2
C_1}}\sqrt[4]{C_3} r\big\}+ \tan ^{-1}\big\{\sqrt{C_3}
r\big\}\\\label{g53a} &\times\xi C_3^{3/4}\big(2 C_1 \sqrt{C_3}-5
C_2\big)-\frac{(3 \xi +8 \pi ) \big(2 C_1 \sqrt{C_3}-C_2\big)
C_3^{5/4} r}{C_3 r^2+1}\bigg]+\frac{A_1}{r},
\end{align}
where $A_1$ is the integration constant with dimension of length and
we take it to be zero to obtain a non-singular solution at $r=0$.
The deformed $g_{rr}$ metric component takes the form as
\begin{equation}\label{g54}
e^{a_2(r)}=\frac{1+C_{3}r^2}{1+\zeta\mathrm{T}^*\big(1+C_{3}r^2\big)}.
\end{equation}
Finally, the effective physical parameters \eqref{g13} are defined
as
\begin{align}\nonumber
\tilde{\mu}&=\frac{\sqrt{C_3} (1-\zeta)}{4 (\xi +2 \pi ) (\xi +4 \pi
) \big(C_2 \sqrt{C_3} r^2+2 C_1\big) \big(C_3 r^2+1\big)^2}\bigg[2
C_1 \sqrt{C_3} \big\{4(\xi +3 \pi )\\\label{g53b}&+(\xi +4 \pi ) C_3
r^2\big\}+C_2 \big\{2 \xi +(\xi +4 \pi ) C_3^2 r^4+6 (\xi +2 \pi )
C_3 r^2\big\}\bigg],\\\nonumber \tilde{P}_{r}&=\frac{-1}{8C_3^{5/4}
r^2 (\xi +2 \pi ) (\xi +4 \pi ) \big(C_2-2 C_1 \sqrt{C_3}\big)
\big(C_3 r^2+1\big)}\bigg[\zeta  r\big(C_3 r^2+1\big) \\\nonumber
&\times\bigg(\frac{4 \sqrt{C_3} C_2}{C_2 \sqrt{C_3} r^2+2
C_1}+\frac{1}{r^2}\bigg) \bigg\{\frac{(3 \xi +8 \pi ) \big(C_2-2 C_1
\sqrt{C_3}\big) C_3^{5/4} r}{C_3 r^2+1}-2 (\xi +4 \pi )\\\nonumber
&\times \big(C_2-2 C_1 \sqrt{C_3}\big) C_3^{5/4} r+4 \sqrt{2} \xi
\sqrt{C_1} \sqrt{C_2} C_3 \tan ^{-1}\bigg(\frac{\sqrt{C_2}
\sqrt[4]{C_3} r}{\sqrt{2} \sqrt{C_1}}\bigg)+C_3^{3/4}\\\nonumber
&\times \xi\big(2 C_1 \sqrt{C_3}-5 C_2\big) \tan
^{-1}\big(\sqrt{C_3} r\big)\bigg\}+2 \big(C_2-2 C_1 \sqrt{C_3}\big)
C_3^{5/4}\bigg\{(\xi +4 \pi ) \\\label{g53c} &\times C_3r^2+\xi
\bigg(\frac{1}{C_3 r^2+1}-7\bigg)+\frac{4 (3 \xi +8 \pi ) C_1}{C_2
\sqrt{C_3} r^2+2 C_1}-16 \pi \bigg\}\bigg],\\\nonumber
\tilde{P}_{\bot}&=\frac{-1}{8C_3^{5/4} r^2 (\xi +2 \pi ) (\xi +4
\pi)\big(C_2-2 C_1 \sqrt{C_3}\big) \big(C_3 r^2+1\big)}\bigg[\zeta r
\big(C_3 r^2+1\big)\\\nonumber &\times \bigg(\frac{4 C_3 C_2^2
r^2}{\big(C_2 \sqrt{C_3} r^2+2 C_1\big)^2}+\frac{2 \sqrt{C_3}
C_2}{C_2 \sqrt{C_3} r^2+2 C_1}+\frac{4 C_1 C_2 \sqrt{C_3}-2 C_2^2
C_3 r^2}{\big(C_3^{3/2} r^2+2 C_1\big)^2}\bigg) \\\nonumber &\times
\bigg\{\frac{(3 \xi +8 \pi ) \big(C_2-2 C_1 \sqrt{C_3}\big)
C_3^{5/4} r}{C_3 r^2+1}+2 (\xi +4 \pi ) \big(2 C_1
\sqrt{C_3}-C_2\big) C_3^{5/4} r\\\nonumber &+4 \sqrt{2} \xi
\sqrt{C_1} \sqrt{C_2} C_3 \tan ^{-1}\bigg(\frac{\sqrt{C_2}
\sqrt[4]{C_3} r}{\sqrt{2} \sqrt{C_1}}\bigg)+\xi  \big(2 C_1
\sqrt{C_3}-5 C_2\big) C_3^{3/4}\\\nonumber &\times \tan
^{-1}\big(\sqrt{C_3} r\big)\bigg\}+2 \big(C_2-2 C_1 \sqrt{C_3}\big)
C_3^{5/4} \bigg\{\xi  \bigg(\frac{1}{C_3 r^2+1}-7\bigg)+ C_3
r^2\\\nonumber &\times(\xi +4 \pi )+\frac{4 (3 \xi +8 \pi ) C_1}{C_2
\sqrt{C_3} r^2+2 C_1}-16 \pi \bigg\}+C_3^{5/4} \zeta
\bigg(\frac{\sqrt{C_3} C_2 r}{C_2 \sqrt{C_3} r^2+2 C_1}+\frac{1}{2
r}\bigg)\\\nonumber &\times \big(C_3 r^2+1\big)\bigg\{C_2
\bigg(\frac{5 \xi  \tan ^{-1}\big(\sqrt{C_3}
r\big)}{\sqrt{C_3}}-\frac{5 \xi  r}{C_3 r^2+1}-\frac{2 (3 \xi +8 \pi
) C_3 r^3}{\big(C_3 r^2+1\big)^2}\bigg)\\\nonumber &+C_1
\bigg(\frac{8 \xi C_2 r}{C_2 \sqrt{C_3} r^2+2 C_1}-2 \xi \tan
^{-1}\big(\sqrt{C_3} r\big)+\frac{4 (3 \xi +8 \pi ) C_3^{3/2}
r^3}{\big(C_3 r^2+1\big)^2}\\\label{g53d} &+\frac{2 \xi \sqrt{C_3}
r}{C_3 r^2+1}\bigg)-\frac{4 \sqrt{2} \xi \sqrt{C_1} \sqrt{C_2} \tan
^{-1}\big(\frac{\sqrt{C_2} \sqrt[4]{C_3} r}{\sqrt{2}
\sqrt{C_1}}\big)}{\sqrt[4]{C_3}}\bigg\}\bigg].
\end{align}
We can obtain the corresponding anisotropy by combining
Eqs.\eqref{g14}, \eqref{g53c} and \eqref{g53d}.

The graphical interpretation for this solution is given by choosing
the estimated data such as the mass $M=1.04 \pm 0.09 M_{\bigodot}$
and radius $R=8.301 \pm 0.2 km$ of a star LMC X-4 \cite{42aa}. As
far the parametric values are concerned, we adopt the decoupling and
model parameters as $\zeta=0.1,0.2,0.3$ and
$\xi=0.25,~0.5,~0.75,~1$, respectively to observe the profile of the
matter sector, deformed metric potential, mass, anisotropy and
energy bounds for the above resulted solution. A large body of
literature exists \cite{42za}-\cite{42ze}, indicating that the
parameter $\xi$ must assume negative values in order to guarantee
hydrostatic equilibrium of stellar configurations. In this regard,
we consider both values of the model parameter to check its impact
on physical properties of the considered compact star. The energy
density remains positive for both positive and negative values of
$\xi$. However, the graph of the radial pressure becomes problematic
because it takes negative values near the spherical boundary. Hence,
we assume only positive values of this parameter to graphically
interpret the solutions. The function \eqref{g53a} is independent of
the decoupling parameter, however, the left plot of Figure
\textbf{1} shows that it possesses increasing trend outwards for all
values of $\xi$. The $g_{rr}$ component \eqref{g54} is also plotted
in the same Figure that starts from $1$ and shows increasing
behavior.
\begin{figure}\center
\epsfig{file=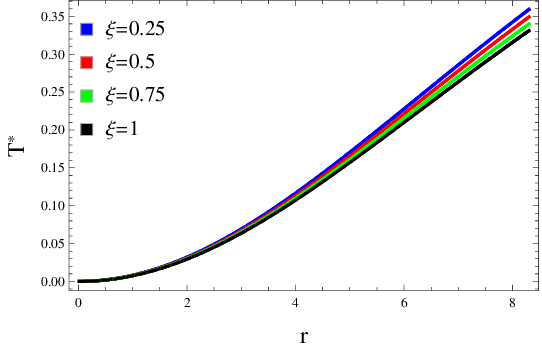,width=0.4\linewidth}\epsfig{file=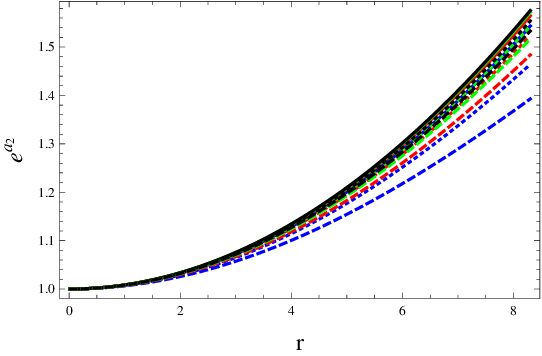,width=0.4\linewidth}
\caption{Deformation function ($\mathrm{T}^*$) \eqref{g53a} and
extended component ($e^{a_2}$) \eqref{g54} for $\zeta=0.1$ (solid),
$0.2$ (dotted) and $0.3$ (dashed) corresponding to model I.}
\end{figure}
\begin{figure}\center
\epsfig{file=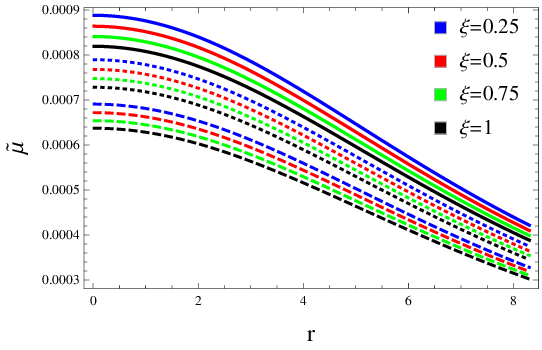,width=0.4\linewidth}\epsfig{file=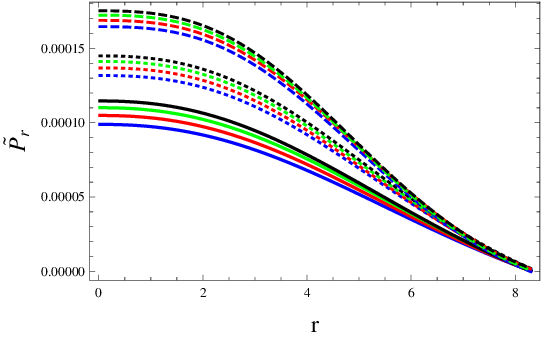,width=0.4\linewidth}
\epsfig{file=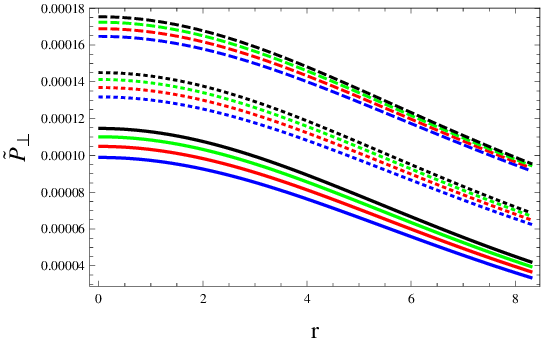,width=0.4\linewidth}\epsfig{file=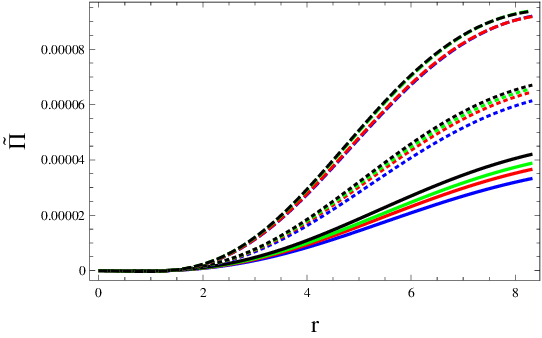,width=0.4\linewidth}
\caption{Effective energy density ($\tilde{\mu}$), radial pressure
($\tilde{P}_r$), tangential pressure ($\tilde{P}_\bot$) and
Anisotropy ($\tilde{\Pi}$) for $\zeta=0.1$ (solid), $0.2$ (dotted)
and $0.3$ (dashed) corresponding to model I.}
\end{figure}

The effective matter determinants \eqref{g53b}-\eqref{g53d} and the
anisotropic factor corresponding to the density-like constraint are
plotted in Figure \textbf{2}. We find that the increment in both
$\zeta$ and $\xi$ results in decrement in the energy density.
However, the radial/transverse pressures exhibit a completely
opposite behavior for these parametric values. The last plot of
Figure \textbf{2} indicates the null anisotropy at the center of a
considered candidate while increasing outwards. We deduce that the
larger values of parameters produce stronger anisotropy in the
self-gravitating interior, i.e., maximum anisotropy at the surface
is achieved for $\zeta=0.3$ and $\xi=1$. Tables
\textbf{1}-\textbf{3} suggest the values of these parameters,
fulfilling the required range for the existence of dense compact
objects. Figure \textbf{3} exhibits the mass function that is found
to be an increasing function from $0<r<R$. Also, the structure
becomes more dense for lower values of parameters. The plots of
compactness and surface redshift indicate lower values than their
proposed upper bounds. Figures \textbf{4} and \textbf{5} display the
dominant energy bounds and stability criteria. It is found that the
model I is viable and stable for every parametric value of $\xi$ and
$\zeta$.
\begin{table}[H]
\scriptsize \centering \caption{Values of some physical parameters
for $\zeta=0.1$ corresponding to Model I.} \label{Table1}
\vspace{+0.07in} \setlength{\tabcolsep}{0.95em}
\begin{tabular}{ccccccc}
% after \\: \hline or \cline{col1-col2} \cline{col3-col4} ...
\hline\hline $\xi$ & 0.25 & 0.5 & 0.75 & 1
\\\hline $\mu_c~(gm/cm^3)$ & 1.1863$\times$10$^{15}$ & 1.1551$\times$10$^{15}$ & 1.1245$\times$10$^{15}$ &
1.0956$\times$10$^{15}$
\\\hline $\mu_s~(gm/cm^3)$ & 5.6551$\times$10$^{14}$ & 5.4865$\times$10$^{14}$ & 5.3393$\times$10$^{14}$ &
5.1841$\times$10$^{14}$
\\\hline $P_c ~(dyne/cm^2)$ & 1.1893$\times$10$^{35}$ & 1.2505$\times$10$^{35}$ & 1.3239$\times$10$^{35}$ &
1.3852$\times$10$^{35}$
\\\hline $\nu_s$ & 0.197 & 0.188 & 0.184 & 0.179
\\\hline
$\mathrm{z}_s$ & 0.279 & 0.267 & 0.258 & 0.247\\
\hline\hline
\end{tabular}
\end{table}
\begin{table}[H]
\scriptsize \centering \caption{Values of some physical parameters
for $\zeta=0.2$ corresponding to Model I.} \label{Table2}
\vspace{+0.07in} \setlength{\tabcolsep}{0.95em}
\begin{tabular}{ccccccc}
% after \\: \hline or \cline{col1-col2} \cline{col3-col4} ...
\hline\hline $\xi$ & 0.25 & 0.5 & 0.75 & 1
\\\hline $\mu_c~(gm/cm^3)$ & 1.0578$\times$10$^{15}$ & 1.0268$\times$10$^{15}$ & 1.0026$\times$10$^{15}$ &
9.7381$\times$10$^{14}$
\\\hline $\mu_s~(gm/cm^3)$ & 5.0477$\times$10$^{14}$ & 4.8818$\times$10$^{14}$ & 4.7467$\times$10$^{14}$ &
4.6704$\times$10$^{14}$
\\\hline $P_c ~(dyne/cm^2)$ & 1.5799$\times$10$^{35}$ & 1.6533$\times$10$^{35}$ & 1.6954$\times$10$^{35}$ &
1.7399$\times$10$^{35}$
\\\hline $\nu_s$ & 0.172 & 0.166 & 0.161 & 0.157
\\\hline
$\mathrm{z}_s$ & 0.232 & 0.227 & 0.217 & 0.212\\
\hline\hline
\end{tabular}
\end{table}
\begin{table}[H]
\scriptsize \centering \caption{Values of some physical parameters
for $\zeta=0.3$ corresponding to Model I.} \label{Table3}
\vspace{+0.07in} \setlength{\tabcolsep}{0.95em}
\begin{tabular}{ccccccc}
% after \\: \hline or \cline{col1-col2} \cline{col3-col4} ...
\hline\hline $\xi$ & 0.25 & 0.5 & 0.75 & 1
\\\hline $\mu_c~(gm/cm^3)$ & 9.2244$\times$10$^{14}$ & 8.9809$\times$10$^{14}$ & 8.7428$\times$10$^{14}$ &
8.4913$\times$10$^{14}$
\\\hline $\mu_s~(gm/cm^3)$ & 4.3895$\times$10$^{14}$ & 4.2691$\times$10$^{14}$ & 4.1767$\times$10$^{14}$ &
4.0429$\times$10$^{14}$
\\\hline $P_c ~(dyne/cm^2)$ & 1.9719$\times$10$^{35}$ & 2.0333$\times$10$^{35}$ & 2.0694$\times$10$^{35}$ &
2.1066$\times$10$^{35}$
\\\hline $\nu_s$ & 0.149 & 0.145 & 0.142 & 0.139
\\\hline
$\mathrm{z}_s$ & 0.197 & 0.192 & 0.181 & 0.175\\
\hline\hline
\end{tabular}
\end{table}
\begin{figure}\center
\epsfig{file=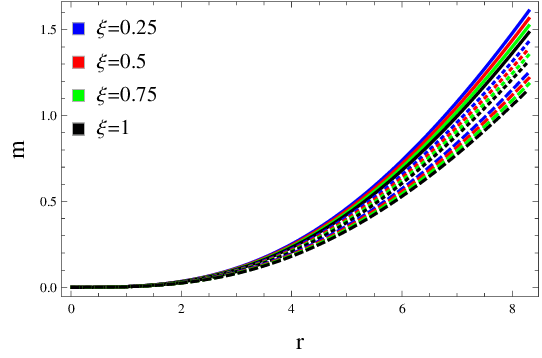,width=0.4\linewidth}\epsfig{file=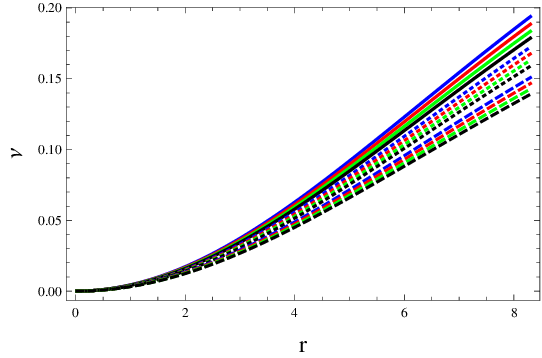,width=0.4\linewidth}
\epsfig{file=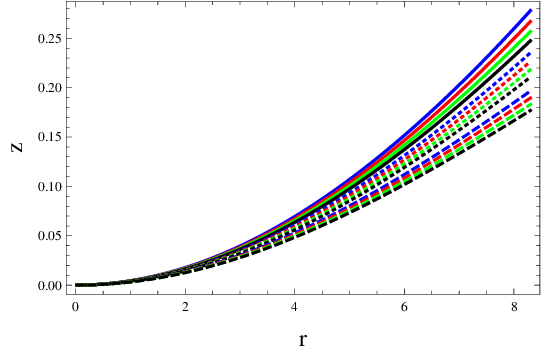,width=0.4\linewidth} \caption{Mass ($m$),
compactness ($\nu$) and surface redshift ($z$) for $\zeta=0.1$
(solid), $0.2$ (dotted) and $0.3$ (dashed) corresponding to model
I.}
\end{figure}
\begin{figure}\center
\epsfig{file=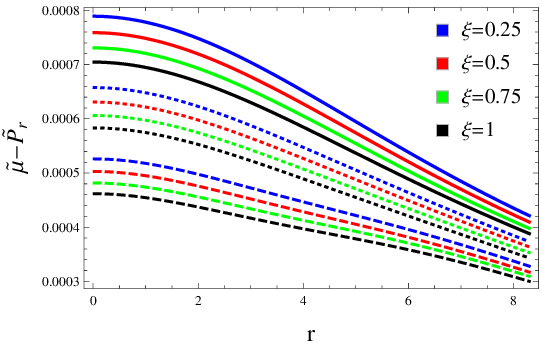,width=0.4\linewidth}\epsfig{file=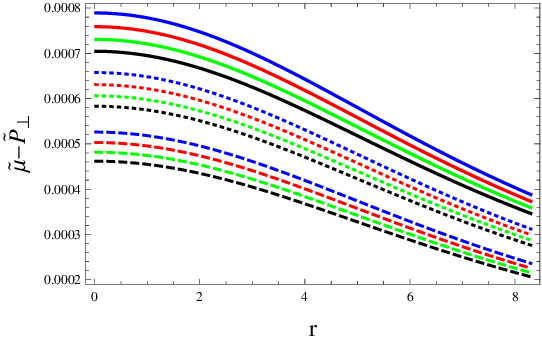,width=0.4\linewidth}
\caption{Energy bounds ($\tilde{\mu}-\tilde{P}_r$ and
$\tilde{\mu}-\tilde{P}_\bot$) for $\zeta=0.1$ (solid), $0.2$
(dotted) and $0.3$ (dashed) corresponding to model I.}
\end{figure}
\begin{figure}\center
\epsfig{file=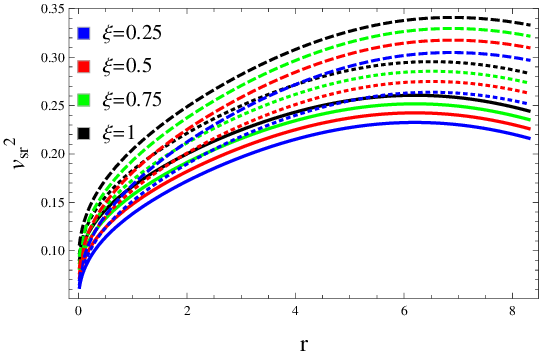,width=0.4\linewidth}\epsfig{file=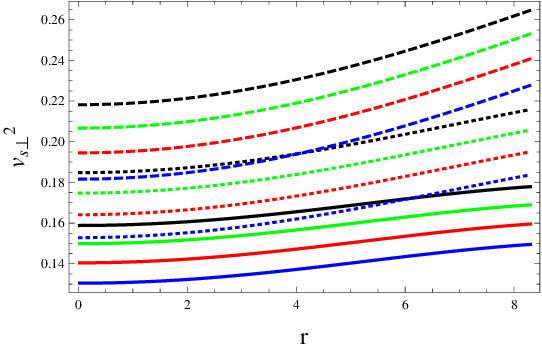,width=0.4\linewidth}
\epsfig{file=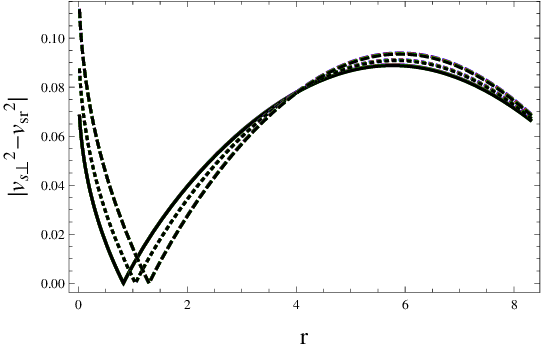,width=0.4\linewidth} \caption{Radial
($v_{sr}^2$), tangential ($v_{s\bot}^2$) sound speeds and cracking
($|v_{s\bot}^2-v_{sr}^2|$) for $\zeta=0.1$ (solid), $0.2$ (dotted)
and $0.3$ (dashed) corresponding to model I.}
\end{figure}

\subsection{Model II}

This subsection adopts a pressure-like constraint \cite{42babb} that
equals the pressures of both parent and additional fluid sources to
construct the solution of the system \eqref{g21}-\eqref{g23}. This
is widely adopted constraint in the literature in order to extend
the solutions from isotropic to the anisotropic domain. The
mathematical expression for this constraint is
\begin{equation}\label{g55}
P=\mathrm{E}_{1}^{1}.
\end{equation}
Equation \eqref{g55} yields after joining with \eqref{g19a} and
\eqref{g23} as
\begin{align}\nonumber
&\frac{\mathrm{T}^*(r)}{8\pi}\bigg\{\frac{a_1'}{r}+\frac{1}{r^2}\bigg\}
-\frac{e^{-a_2}}{8r^2\big(\xi^2+6\pi\xi+8\pi^2\big)}\\\label{g56}
&\times\big[(3 \xi +8 \pi ) r a_1 '+\xi  r a_2 '-2 (\xi +4 \pi )
\big(e^{a_2 }-1\big)\big]=0,
\end{align}
that becomes in relation with the metric \eqref{g33}
\begin{align}\nonumber
&\bigg(\frac{4 \sqrt{C_3} C_2}{C_2 \sqrt{C_3} r^2+2
C_1}+\frac{1}{r^2}\bigg) \mathrm{T}^*(r)+\frac{2 \pi}{(\xi +2 \pi )
(\xi +4 \pi ) r^2 \big(C_3 r^2+1\big)}\\\label{g57}
&\times\bigg\{(\xi +4 \pi ) C_3 r^2+\xi \bigg(\frac{1}{C_3
r^2+1}-7\bigg)+\frac{4 (3 \xi +8 \pi ) C_1}{C_2 \sqrt{C_3} r^2+2
C_1}-16 \pi \bigg\}=0,
\end{align}
providing the deformation function as follows
\begin{align}\nonumber
\mathrm{T}^*(r)&=\frac{2 \pi  \sqrt{C_3} r^2}{(\xi +2 \pi ) (\xi +4
\pi ) \left(5 C_2 \sqrt{C_3} r^2+2 C_1\right) \left(C_3
r^2+1\right)^2}\bigg[C_2 \big\{2 (3 \xi +8 \pi )\\\label{g58}&-(\xi
+4 \pi ) C_3^2 r^4+6 (\xi +2 \pi ) C_3 r^2\big\}-2 C_1 \sqrt{C_3}
\big\{(\xi +4 \pi ) C_3 r^2+4 \pi \big\}\bigg].
\end{align}
The deformed $g_{rr}$ component for the constraint \eqref{g55}
becomes
\begin{align}\nonumber
e^{a_2(r)}&=\big[(\xi +2 \pi ) (\xi +4 \pi ) \big(5 C_2 \sqrt{C_3}
r^2+2 C_1\big) \big(C_3 r^2+1\big)^2\big]\big[(\xi +2 \pi )(\xi +4
\pi )\\\nonumber &\times \big(5 C_2 \sqrt{C_3} r^2+2 C_1\big)
\big(C_3 r^2+1\big)+2 \pi \sqrt{C_3} \zeta  r^2 \big\{C_2 \big(6
\xi-(\xi +4 \pi ) C_3^2 r^4\\\label{g58a} & +6 (\xi +2 \pi ) C_3
r^2+16 \pi \big)-2 C_1 \sqrt{C_3} \big((\xi +4 \pi ) C_3 r^2+4 \pi
\big)\big\}\big]^{-1}.
\end{align}
Hence, the effective energy density, pressure components \eqref{g13}
and anisotropy \eqref{g14} are given by
\begin{align}\nonumber
\tilde{\mu}&=\frac{\sqrt{C_3}}{4 (\xi +2 \pi ) (\xi +4 \pi )
\big(C_2 \sqrt{C_3} r^2+2 C_1\big) \big(5 C_2 \sqrt{C_3} r^2+2
C_1\big)^2 \big(C_3 r^2+1\big)^3}\\\nonumber &\times\bigg[2 C_1
C_2^2 \sqrt{C_3} r^2 \big\{20 \xi-16 (3 \xi +8 \pi ) \zeta +C_3 r^2
\big(12 (\xi  (15-2 \zeta )+(3 \zeta +35)\\\nonumber &\times
\pi)+C_3 r^2 \big(\xi (71 \zeta +195)+(\xi +4 \pi ) C_3 (3 \zeta
+35) r^2+16 \pi (11 \zeta +35)\big)\big)\big\}\\\nonumber &+8 C_1^3
\sqrt{C_3} \big\{4 (\xi +3 \pi (\zeta +1))+C_3 (\zeta +1) r^2 \big(5
\xi +(\xi +4 \pi ) C_3 r^2+16 \pi \big)\big\}\\\nonumber &-4 C_1^2
C_2 \big\{2 \xi (9 \zeta -1)+C_3 r^2 \big(12 (2 \xi (\zeta -2)+\pi
(\zeta -11))+C_3 r^2 \big(-3 \xi \\\nonumber &\times(7 \zeta
+19)+(\xi +4 \pi ) C_3 (\zeta -11) r^2-16 \pi (2 \zeta
+11)\big)\big)+48 \pi \zeta \big\}+5 C_2^3 C_3 \\\nonumber &\times
r^4 \big\{C_3 r^2 \big(40 \xi +C_3 r^2 \big(\xi (11 \zeta +35)+(\xi
+4 \pi ) C_3 (\zeta +5) r^2+16 \pi (2 \zeta +5)\big)\\\label{g58b}
&+12 \pi (\zeta +5)\big)-2 (\xi (3 \zeta -5)+8 \pi \zeta
)\big\}\bigg],\\\nonumber \tilde{P}_{r}&=\frac{-\sqrt{C_3} (\zeta
+1)}{4 (\xi +2 \pi ) (\xi +4 \pi ) \big(C_2 \sqrt{C_3} r^2+2
C_1\big) \big(C_3 r^2+1\big)^2}\bigg[2 C_1 \sqrt{C_3} \big\{4 \pi+
C_3 r^2\\\label{g58c} &\times(\xi +4 \pi ) \big\}+C_2 \big\{(\xi +4
\pi ) C_3^2 r^4-2 (3 \xi +8 \pi )-6 (\xi +2 \pi ) C_3
r^2\big\}\bigg],\\\nonumber \tilde{P}_{\bot}&=\frac{-\sqrt{C_3}}{4
(\xi +2 \pi ) (\xi +4 \pi ) \big(C_2 \sqrt{C_3} r^2+2 C_1\big)
\big(5 C_2 \sqrt{C_3} r^2+2 C_1\big)^2 \big(C_3
r^2+1\big)^3}\\\nonumber &\times\bigg[2 C_1 C_2^2 \sqrt{C_3} r^2
\big\{C_3 r^2 \big(C_3 r^2 \big(64 \xi \zeta -25 \xi +(\xi +4 \pi )
C_3 (12 \zeta +35) r^2\\\nonumber &+204 \pi \zeta +120 \pi \big)-4
(9 \xi  \zeta +30 \xi -11 \pi  \zeta +45 \pi )\big)-(3 \xi +8 \pi )
(7 \zeta +10)\\\nonumber &\times 2\big\}+8 C_1^3 \sqrt{C_3}
\big\{C_3 r^2 \big(2 \xi  \zeta +\xi +(\xi +4 \pi ) C_3 r^2+4 \pi
(\zeta +2)\big)+ \pi (\zeta +1)\\\nonumber &\times 4\big\}+4 C_1^2
C_2 \big\{C_3 r^2 \big(2 (2 \pi  (5 \zeta +3)-3 \xi (\zeta +2))+C_3
r^2 \big(\xi (18 \zeta +5)+11 \\\nonumber &\times(\xi +4 \pi ) C_3
r^2+36 \pi (\zeta +2)\big)\big)-2 (3 \xi +8 \pi ) (\zeta +1)\big\}+5
C_2^3 C_3 r^4\big\{C_3 r^2 \\\nonumber &\times \big(C_3 r^2 \big(4
(\xi +7 \pi ) \zeta -5 (5 \xi +8 \pi )+(\xi +4 \pi ) C_3 (4 \zeta
+5) r^2\big)-2 ( (\zeta +2)\\\label{g58d} &\times15 \xi+\pi (26
\zeta +70))\big)-2 (3 \xi +8 \pi ) (4 \zeta
+5)\big\}\bigg],\\\nonumber \tilde{\Pi}&=\frac{\zeta r^2 C_3}{4 (\xi
+2 \pi ) (\xi +4 \pi ) \big(C_2 \sqrt{C_3} r^2+2 C_1\big) \big(5 C_2
\sqrt{C_3} r^2+2 C_1\big)^2 \big(C_3 r^2+1\big)^3}\\\nonumber
&\times\bigg[8 C_3 C_1^3 \big((\xi +4 \pi ) C_3 r^2-\xi+4 \pi
\big)+4 C_2 \sqrt{C_3} C_1^2 \big(C_3 r^2 \big(11 (\xi +4 \pi ) C_3
r^2\\\nonumber &+36 \pi-13 \xi \big)-6 \xi-8 \pi \big)+2 C_2^2 C_1
\big(C_3 r^2 \big(C_3 r^2 \big(23 (\xi +4 \pi ) C_3 r^2-89
\xi\\\nonumber &-84 \pi \big)-28 (3 \xi +8 \pi )\big)-6 (3 \xi +8
\pi )\big)+5 C_2^3 \sqrt{C_3} r^2 \big(C_3 r^2 \big(C_3 r^2 \big(C_3
r^2\\\label{g58e} &\times(\xi +4 \pi ) -29 \xi-68 \pi \big)-30
\xi-88 \pi \big)-2 (3 \xi +8 \pi )\big)\bigg].
\end{align}

The behavior of the deformation function \eqref{g58} is shown in the
left plot of Figure \textbf{6} that becomes null at the center,
increases up to $r\thickapprox 5.5$ and then again decreases towards
the boundary. The same Figure also exhibits the extended radial
component \eqref{g58a}, presenting an acceptable (increasing)
behavior throughout. The desired behavior (already discussed in
section 5) of the effective energy density and both pressure
components \eqref{g58b}-\eqref{g58d} are observed in Figure
\textbf{7}, thus they reign the interior of a compact object. The
behavior of such effective parameters in relation with $\xi$ and
$\zeta$ are found the same as for the model I. Tables
\textbf{4}-\textbf{6} present their values for all choices of these
parameters which are in accordance with the observed data for
compact stars. The vanishing radial pressure at $r=R=8.301$ confirms
the continuity of the second fundamental form of the matching
conditions. Moreover, there does not exist pressure anisotropy at
the center that initially decreases outwards and then again
increases towards the spherical junction (lower right plot of Figure
\textbf{7}).
\begin{table}[h!]
\scriptsize \centering \caption{Values of some physical parameters
for $\zeta=0.1$ corresponding to Model II.} \label{Table4}
\vspace{+0.07in} \setlength{\tabcolsep}{0.95em}
\begin{tabular}{ccccccc}
% after \\: \hline or \cline{col1-col2} \cline{col3-col4} ...
\hline\hline $\xi$ & 0.25 & 0.5 & 0.75 & 1
\\\hline $\mu_c~(gm/cm^3)$ & 1.2897$\times$10$^{15}$ & 1.2542$\times$10$^{15}$ & 1.2177$\times$10$^{15}$ &
1.1832$\times$10$^{15}$
\\\hline $\mu_s~(gm/cm^3)$ & 6.2571$\times$10$^{14}$ & 6.1006$\times$10$^{14}$ & 5.9253$\times$10$^{14}$ &
5.7701$\times$10$^{14}$
\\\hline $P_c ~(dyne/cm^2)$ & 8.7187$\times$10$^{34}$ & 9.6266$\times$10$^{34}$ & 1.0427$\times$10$^{35}$ &
1.1133$\times$10$^{35}$
\\\hline $\nu_s$ & 0.188 & 0.184 & 0.178 & 0.173
\\\hline
$\mathrm{z}_s$ & 0.267 & 0.256 & 0.246 & 0.238\\
\hline\hline
\end{tabular}
\end{table}
\begin{table}[h!]
\scriptsize \centering \caption{Values of some physical parameters
for $\zeta=0.2$ corresponding to Model II.} \label{Table5}
\vspace{+0.07in} \setlength{\tabcolsep}{0.95em}
\begin{tabular}{ccccccc}
% after \\: \hline or \cline{col1-col2} \cline{col3-col4} ...
\hline\hline $\xi$ & 0.25 & 0.5 & 0.75 & 1
\\\hline $\mu_c~(gm/cm^3)$ & 1.2669$\times$10$^{15}$ & 1.2249$\times$10$^{15}$ & 1.1868$\times$10$^{15}$ &
1.1496$\times$10$^{15}$
\\\hline $\mu_s~(gm/cm^3)$ & 6.4149$\times$10$^{14}$ & 6.2624$\times$10$^{14}$ & 6.0349$\times$10$^{14}$ &
5.9052$\times$10$^{14}$
\\\hline $P_c ~(dyne/cm^2)$ & 9.5087$\times$10$^{34}$ & 1.0501$\times$10$^{35}$ & 1.1376$\times$10$^{35}$ &
1.2144$\times$10$^{35}$
\\\hline $\nu_s$ & 0.188 & 0.182 & 0.176 & 0.172
\\\hline
$\mathrm{z}_s$ & 0.266 & 0.255 & 0.245 & 0.236\\
\hline\hline
\end{tabular}
\end{table}
\begin{table}[h!]
\scriptsize \centering \caption{Values of some physical parameters
for $\zeta=0.3$ corresponding to Model II.} \label{Table6}
\vspace{+0.07in} \setlength{\tabcolsep}{0.95em}
\begin{tabular}{ccccccc}
% after \\: \hline or \cline{col1-col2} \cline{col3-col4} ...
\hline\hline $\xi$ & 0.25 & 0.5 & 0.75 & 1
\\\hline $\mu_c~(gm/cm^3)$ & 1.2404$\times$10$^{15}$ & 1.1941$\times$10$^{15}$ & 1.1544$\times$10$^{15}$ &
1.1159$\times$10$^{15}$
\\\hline $\mu_s~(gm/cm^3)$ & 6.3909$\times$10$^{14}$ & 6.1728$\times$10$^{14}$ & 6.0082$\times$10$^{14}$ &
5.8397$\times$10$^{14}$
\\\hline $P_c ~(dyne/cm^2)$ & 1.0305$\times$10$^{35}$ & 1.1376$\times$10$^{35}$ & 1.2325$\times$10$^{35}$ &
1.3239$\times$10$^{35}$
\\\hline $\nu_s$ & 0.187 & 0.183 & 0.177 & 0.171
\\\hline
$\mathrm{z}_s$ & 0.267 & 0.255 & 0.243 & 0.235\\
\hline\hline
\end{tabular}
\end{table}
\begin{figure}[H]\center
\epsfig{file=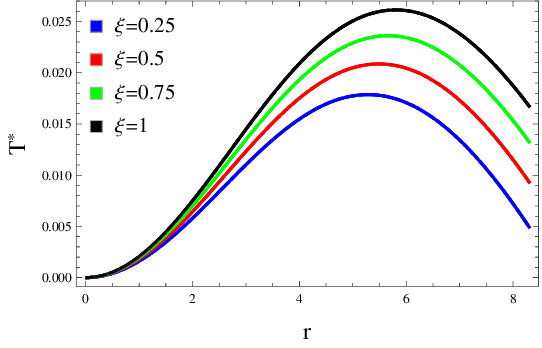,width=0.4\linewidth}\epsfig{file=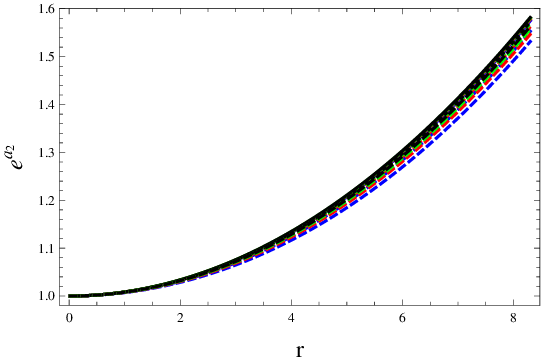,width=0.4\linewidth}
\caption{Deformation function ($\mathrm{T}^*$) \eqref{g58} and
extended component ($e^{a_2}$) \eqref{g58a} for $\zeta=0.1$ (solid),
$0.2$ (dotted) and $0.3$ (dashed) corresponding to model II.}
\end{figure}
\begin{figure}[H]\center
\epsfig{file=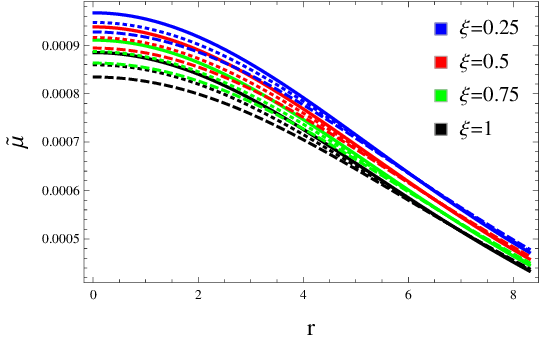,width=0.4\linewidth}\epsfig{file=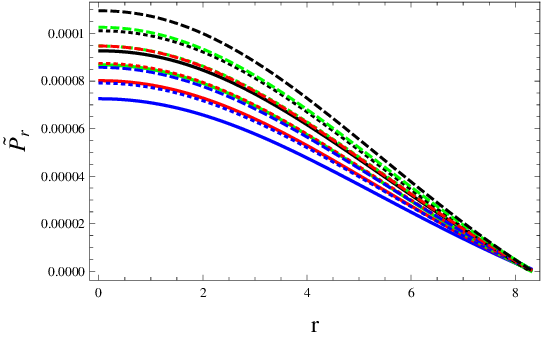,width=0.4\linewidth}
\epsfig{file=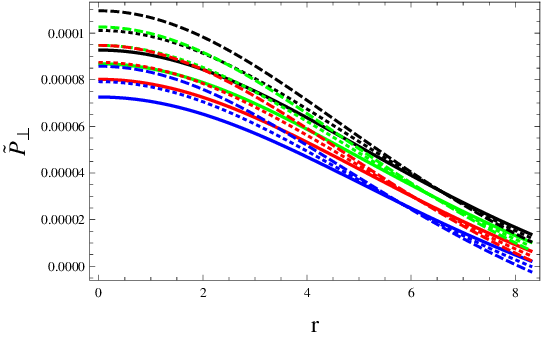,width=0.4\linewidth}\epsfig{file=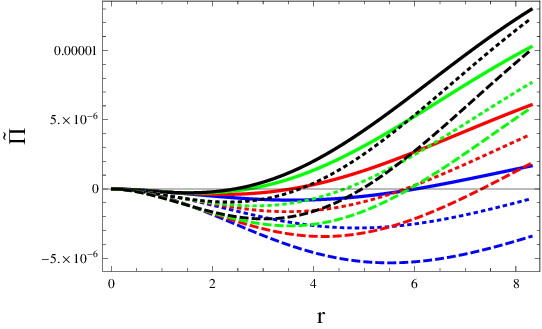,width=0.4\linewidth}
\caption{Effective energy density ($\tilde{\mu}$), radial pressure
($\tilde{P}_r$), tangential pressure ($\tilde{P}_\bot$) and
Anisotropy ($\tilde{\Pi}$) for $\zeta=0.1$ (solid), $0.2$ (dotted)
and $0.3$ (dashed) corresponding to model II.}
\end{figure}

The calculated mass of a star LMC X-4 for our developed model II is
exhibited in Figure \textbf{8} which is consistent with the observed
mass. The other two plots of this Figure indicate an admissible
behavior of the surface redshift and compactness. Two energy bounds
such as $\tilde{\mu}-\tilde{P}_r \geq 0$ and
$\tilde{\mu}-\tilde{P}_\bot \geq 0$ are pictured in Figure
\textbf{9} whose acceptable profile shows the viability of our
solution. Hence, the usual matter exists in the proposed interior.
Finally, Figure \textbf{10} tests the stability criteria and all
three plots confirm that our model II is stable for all parametric
choices.
\begin{figure}[H]\center
\epsfig{file=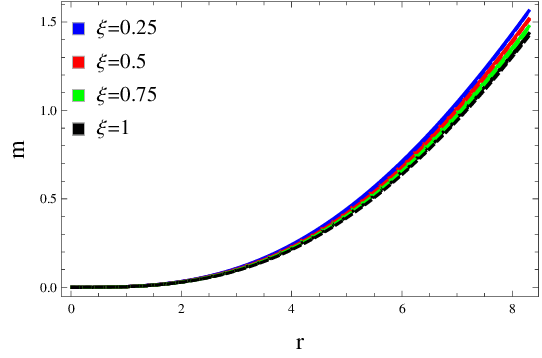,width=0.4\linewidth}\epsfig{file=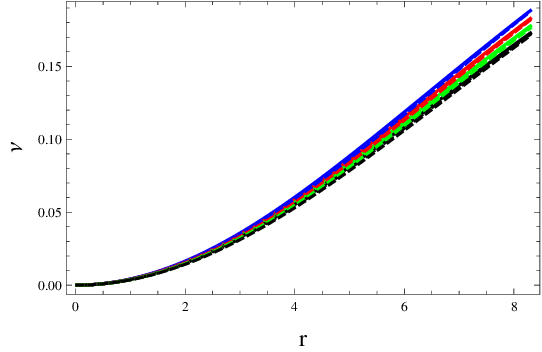,width=0.4\linewidth}
\epsfig{file=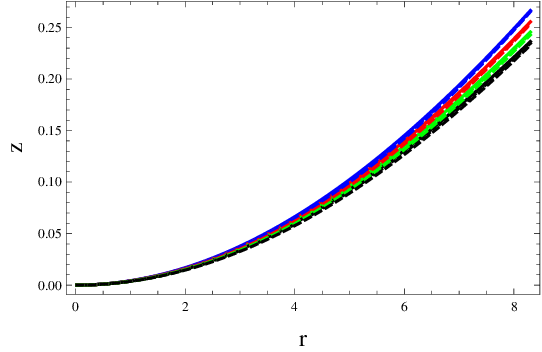,width=0.4\linewidth} \caption{Mass ($m$),
compactness ($\nu$) and surface redshift ($z$) for $\zeta=0.1$
(solid), $0.2$ (dotted) and $0.3$ (dashed) corresponding to model
II.}
\end{figure}
\begin{figure}[H]\center
\epsfig{file=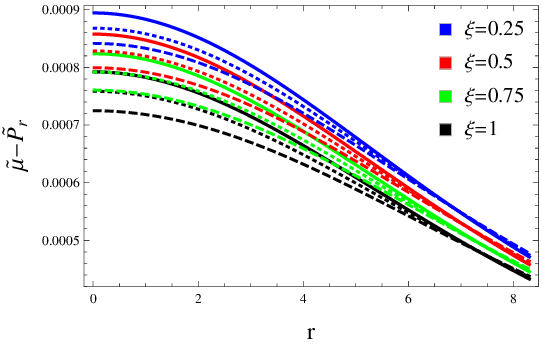,width=0.4\linewidth}\epsfig{file=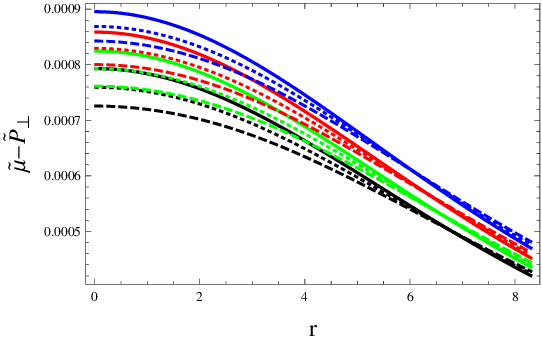,width=0.4\linewidth}
\caption{Energy bounds ($\tilde{\mu}-\tilde{P}_r$ and
$\tilde{\mu}-\tilde{P}_\bot$) for $\zeta=0.1$ (solid), $0.2$
(dotted) and $0.3$ (dashed) corresponding to model II.}
\end{figure}
\begin{figure}[H]\center
\epsfig{file=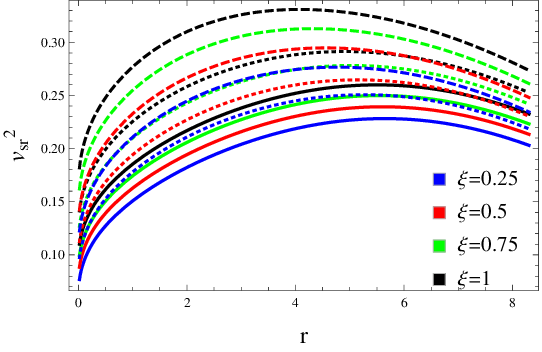,width=0.4\linewidth}\epsfig{file=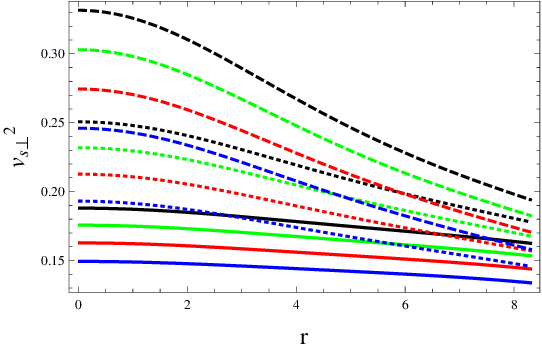,width=0.4\linewidth}
\epsfig{file=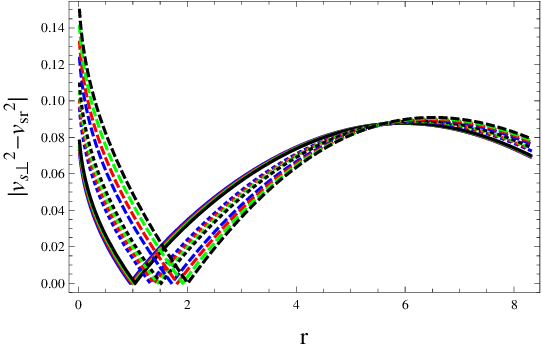,width=0.4\linewidth} \caption{Radial
($v_{sr}^2$), tangential ($v_{s\bot}^2$) sound speeds and cracking
($|v_{s\bot}^2-v_{sr}^2|$) for $\zeta=0.1$ (solid), $0.2$ (dotted)
and $0.3$ (dashed) corresponding to model II.}
\end{figure}

\subsection{Model III}

Here, we adopt a linear equation of state \cite{42babc} that
interlinks the matter sector of the additional fluid source as
another constraint so that a unique solution of the system
\eqref{g21}-\eqref{g23} is obtained. This equation of state is given
as follows
\begin{equation}\label{g59}
\mathrm{E}_{1}^{1}(r)=\kappa_1\mathrm{E}_{0}^{0}(r)+\kappa_2,
\end{equation}
where $\kappa_1$ and $\kappa_2$ are arbitrary constants whose
different values may intensely affect the calculated solution.
Joining Eqs.\eqref{g22}, \eqref{g23} and \eqref{g59} together, we
get the following linear-order different equation
\begin{equation}\label{g60}
\frac{\kappa_1\mathrm{T}^{*'}(r)}{r}-\left(\frac{a_1'}{r}+\frac{1}{r^2}-\frac{\kappa_1}{r^2}\right)\mathrm{T}^*(r)+8\pi\kappa_2=0.
\end{equation}
Note that the third component of an extra fluid configuration
($\mathrm{E}_{2}^{2}$) can also be involved in the constraint
\eqref{g59}, however, the resulting second-order differential
equation becomes more complicated and may not be solved. When we
substitute the metric \eqref{g33} in \eqref{g60}, the differential
equation takes the form
\begin{align}\label{g61}
&\mathrm{T}^*(r) \bigg\{\frac{4 \sqrt{C_3} C_2}{C_2 \sqrt{C_3} r^2+2
C_1}-\frac{\kappa _1}{r^2}+\frac{1}{r^2}\bigg\}-\frac{\kappa _1
\mathrm{T}^{*'}(r)}{r}-8 \pi  \kappa _2=0.
\end{align}

Since the above equation is linear in $\mathrm{T}^*(r)$, we firstly
solve it through exact integration and obtain the deformation
function in terms of hypergeometric terms. We face difficulty while
plotting the corresponding physical factors due to such complex
terms. We, therefore, apply the numerical integration to solve
Eq.\eqref{g61} for $\mathrm{T}^*(r)$ by providing the initial
condition $\mathrm{T}^*(r)|_{r=0}=0$ along with $\kappa_1=1.1$ and
$\kappa_2=0.01$. Figure \textbf{11} shows the variation of the
deformation function with respect to $\xi$ that possesses an
increasing profile for $0<r<R$. The right plot of the same Figure
indicates that the impact of $\xi$ on the extended $g_{rr}$
component is negligible due to the constraint \eqref{g59} that
solely depends on the additional sector. However, it exhibits
increasing trend for every value of $\zeta$. Further, the matter
triplet and anisotropy are obtained through Eqs.\eqref{g13} and
\eqref{g14} and are plotted in Figure \textbf{12}. We observe that
the energy density and both pressures show opposite trend with
respect to $\zeta$ and $\xi$ at the center.
\begin{figure}\center
\epsfig{file=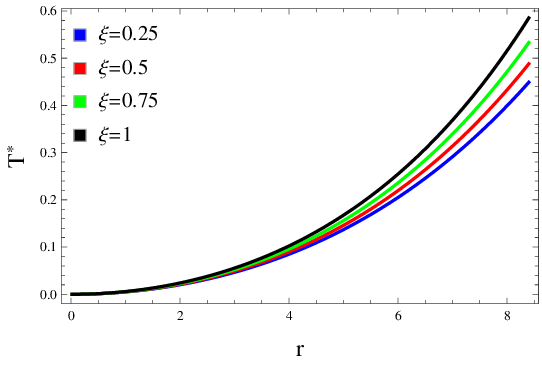,width=0.4\linewidth}\epsfig{file=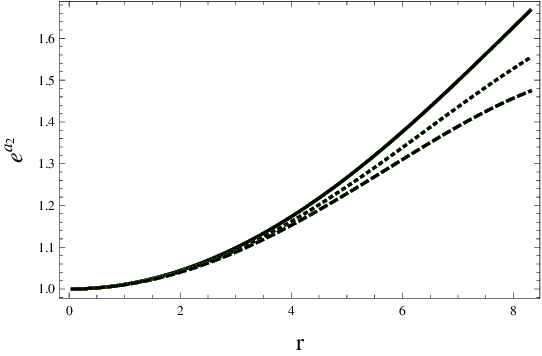,width=0.4\linewidth}
\caption{Deformation function ($\mathrm{T}^*$) and extended
component ($e^{a_2}$) for $\zeta=0.1$ (solid), $0.2$ (dotted) and
$0.3$ (dashed) corresponding to model III.}
\end{figure}
\begin{figure}\center
\epsfig{file=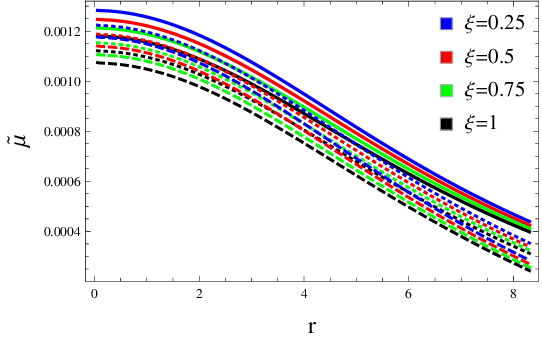,width=0.4\linewidth}\epsfig{file=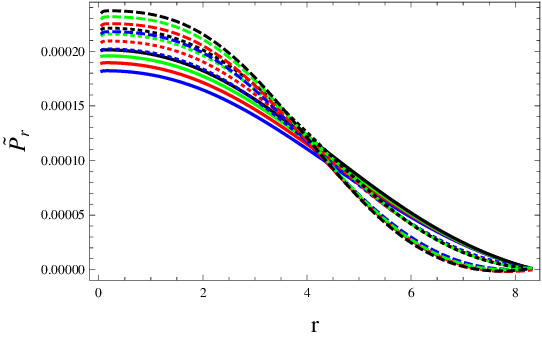,width=0.4\linewidth}
\epsfig{file=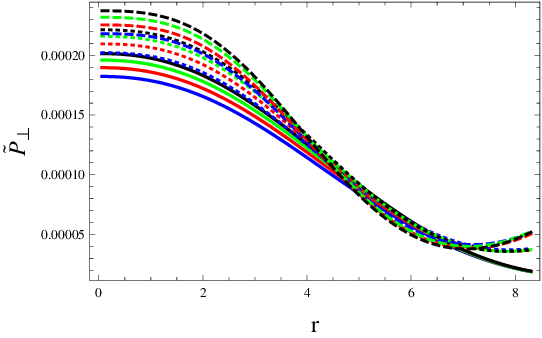,width=0.4\linewidth}\epsfig{file=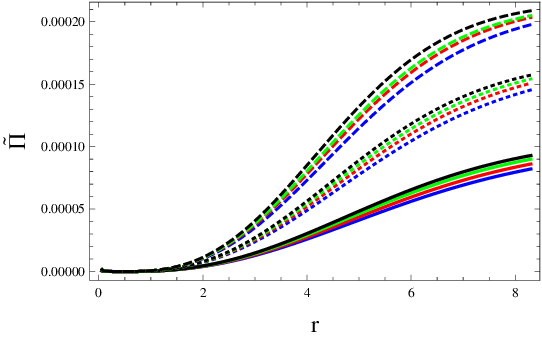,width=0.4\linewidth}
\caption{Effective energy density ($\tilde{\mu}$), radial pressure
($\tilde{P}_r$), tangential pressure ($\tilde{P}_\bot$) and
Anisotropy ($\tilde{\Pi}$) for $\zeta=0.1$ (solid), $0.2$ (dotted)
and $0.3$ (dashed) corresponding to model III.}
\end{figure}

Tables \textbf{7}-\textbf{9} present the numerical values of these
physical quantities, clarifying the effect of both parameters. The
increasing trend of anisotropic factor provides an outward force
that may help to maintain the stability of our model III. Figure
\textbf{13} shows that the mass function is increasing outwards and
the proposed model is more dense for $\xi=0.25$ and $\zeta=0.1$. The
other two factors in this Figure are found within the allowable
limits. Figure \textbf{14} demonstrates that there must exist usual
matter in the interior because the dominant energy conditions are
fulfilled. The stability analysis is shown in Figure \textbf{15}
from which we deduce that our model III is stable only for
$\zeta=0.1$ and $0.2$. However, the tangential sound speed criterion
is not satisfied for $\zeta=0.3$.
\begin{table}
\scriptsize \centering \caption{Values of some physical parameters
for $\zeta=0.1$ corresponding to Model III.} \label{Table7}
\vspace{+0.07in} \setlength{\tabcolsep}{0.95em}
\begin{tabular}{ccccccc}
% after \\: \hline or \cline{col1-col2} \cline{col3-col4} ...
\hline\hline $\xi$ & 0.25 & 0.5 & 0.75 & 1
\\\hline $\mu_c~(gm/cm^3)$ & 1.7164$\times$10$^{15}$ & 1.6669$\times$10$^{15}$ & 1.6215$\times$10$^{15}$ &
1.5746$\times$10$^{15}$
\\\hline $\mu_s~(gm/cm^3)$ & 5.9106$\times$10$^{14}$ & 5.6898$\times$10$^{14}$ & 5.4196$\times$10$^{14}$ &
5.2992$\times$10$^{14}$
\\\hline $P_c ~(dyne/cm^2)$ & 2.1884$\times$10$^{35}$ & 2.2629$\times$10$^{35}$ & 2.3639$\times$10$^{35}$ &
2.4169$\times$10$^{35}$
\\\hline $\nu_s$ & 0.195 & 0.189 & 0.183 & 0.178
\\\hline
$\mathrm{z}_s$ & 0.279 & 0.267 & 0.255 & 0.246\\
\hline\hline
\end{tabular}
\end{table}
\begin{table}
\scriptsize \centering \caption{Values of some physical parameters
for $\zeta=0.2$ corresponding to Model III.} \label{Table8}
\vspace{+0.07in} \setlength{\tabcolsep}{0.95em}
\begin{tabular}{ccccccc}
% after \\: \hline or \cline{col1-col2} \cline{col3-col4} ...
\hline\hline $\xi$ & 0.25 & 0.5 & 0.75 & 1
\\\hline $\mu_c~(gm/cm^3)$ & 1.6375$\times$10$^{15}$ & 1.5853$\times$10$^{15}$ & 1.5425$\times$10$^{15}$ &
1.4944$\times$10$^{15}$
\\\hline $\mu_s~(gm/cm^3)$ & 4.7145$\times$10$^{14}$ & 4.5152$\times$10$^{14}$ & 4.3547$\times$10$^{14}$ &
4.0978$\times$10$^{14}$
\\\hline $P_c ~(dyne/cm^2)$ & 2.4277$\times$10$^{35}$ & 2.5143$\times$10$^{35}$ & 2.5901$\times$10$^{35}$ &
2.6561$\times$10$^{35}$
\\\hline $\nu_s$ & 0.171 & 0.165 & 0.161 & 0.154
\\\hline
$\mathrm{z}_s$ & 0.235 & 0.224 & 0.214 & 0.204\\
\hline\hline
\end{tabular}
\end{table}
\begin{table}[H]
\scriptsize \centering \caption{Values of some physical parameters
for $\zeta=0.3$ corresponding to Model III.} \label{Table9}
\vspace{+0.07in} \setlength{\tabcolsep}{0.95em}
\begin{tabular}{ccccccc}
% after \\: \hline or \cline{col1-col2} \cline{col3-col4} ...
\hline\hline $\xi$ & 0.25 & 0.5 & 0.75 & 1
\\\hline $\mu_c~(gm/cm^3)$ & 1.5719$\times$10$^{15}$ & 1.5251$\times$10$^{15}$ & 1.4797$\times$10$^{15}$ &
1.4302$\times$10$^{15}$
\\\hline $\mu_s~(gm/cm^3)$ & 3.8169$\times$10$^{14}$ & 3.6162$\times$10$^{14}$ & 3.4168$\times$10$^{14}$ &
3.1961$\times$10$^{14}$
\\\hline $P_c ~(dyne/cm^2)$ & 2.6177$\times$10$^{35}$ & 2.7054$\times$10$^{35}$ & 2.7981$\times$10$^{35}$ &
2.8809$\times$10$^{35}$
\\\hline $\nu_s$ & 0.153 & 0.148 & 0.142 & 0.137
\\\hline
$\mathrm{z}_s$ & 0.203 & 0.192 & 0.183 & 0.175\\
\hline\hline
\end{tabular}
\end{table}
\begin{figure}[H]\center
\epsfig{file=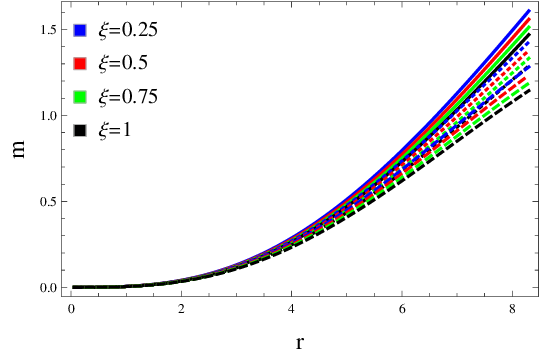,width=0.4\linewidth}\epsfig{file=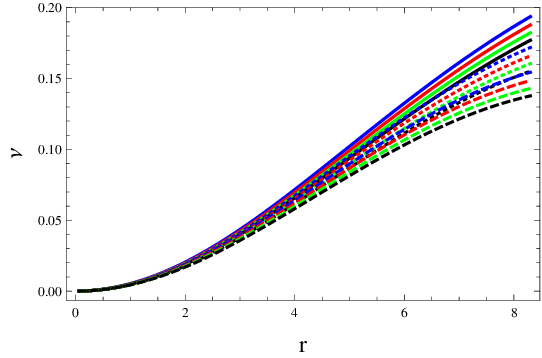,width=0.4\linewidth}
\epsfig{file=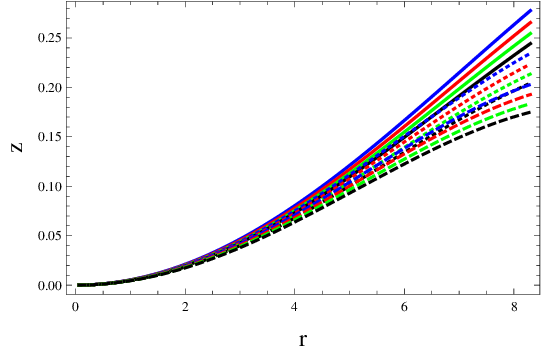,width=0.4\linewidth} \caption{Mass ($m$),
compactness ($\nu$) and surface redshift ($z$) for $\zeta=0.1$
(solid), $0.2$ (dotted) and $0.3$ (dashed) corresponding to model
III.}
\end{figure}
\begin{figure}[H]\center
\epsfig{file=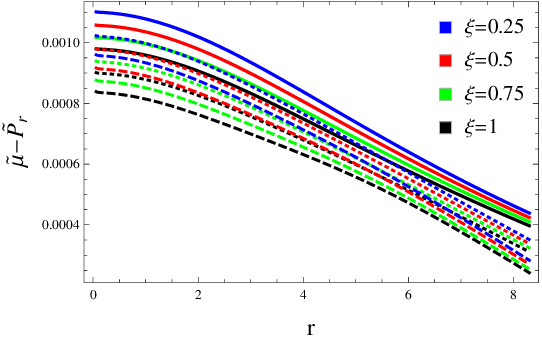,width=0.4\linewidth}\epsfig{file=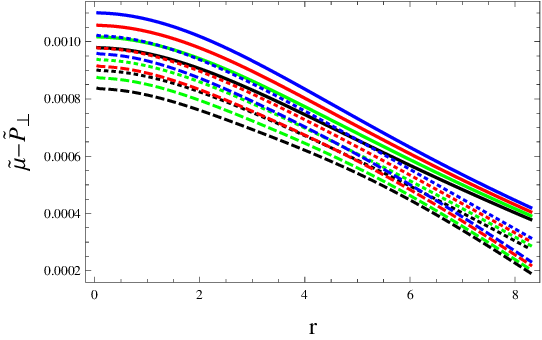,width=0.4\linewidth}
\caption{Energy bounds ($\tilde{\mu}-\tilde{P}_r$ and
$\tilde{\mu}-\tilde{P}_\bot$) for $\zeta=0.1$ (solid), $0.2$
(dotted) and $0.3$ (dashed) corresponding to model III.}
\end{figure}
\begin{figure}[H]\center
\epsfig{file=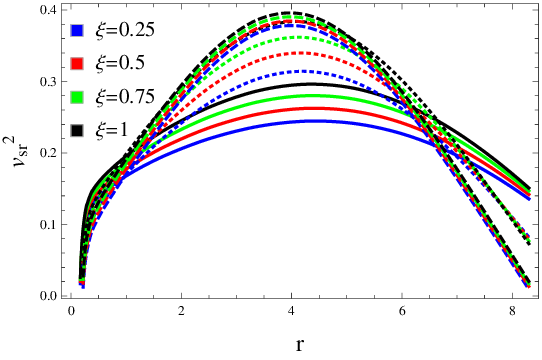,width=0.4\linewidth}\epsfig{file=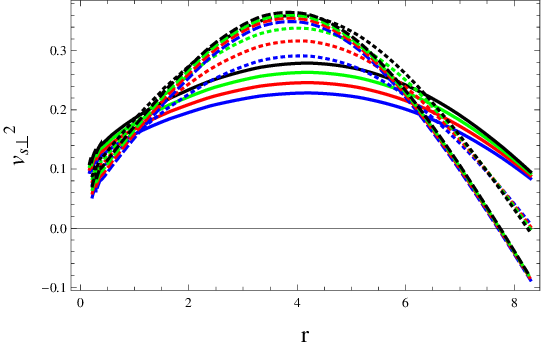,width=0.4\linewidth}
\epsfig{file=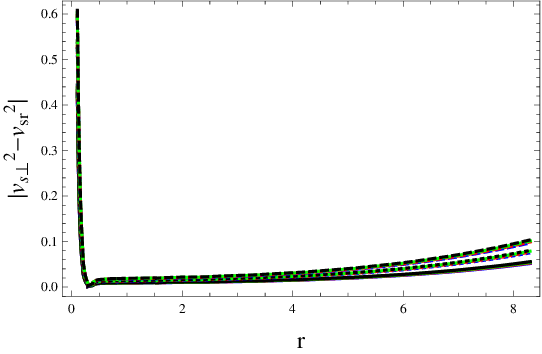,width=0.4\linewidth} \caption{Radial
($v_{sr}^2$), tangential ($v_{s\bot}^2$) sound speeds and cracking
($|v_{s\bot}^2-v_{sr}^2|$) for $\zeta=0.1$ (solid), $0.2$ (dotted)
and $0.3$ (dashed) corresponding to model III.}
\end{figure}

\section{Conclusions}

This paper is devoted to the formulation of three different
extensions of an isotropic Finch-Skea metric to the anisotropic
domain with the help of the MGD technique in a modified linear
$f(\mathcal{R},\mathcal{T})=\mathcal{R}+2\xi\mathcal{T}$ theory. In
order to address this, a static spherical geometry has been assumed
that characterizes the interior self-gravitating fluid distribution.
We have considered that the sphere is originally filled with an
isotropic matter and then added a Lagrangian density analogous to an
extra fluid source, generating anisotropy in the original perfect
fluid. We have added another source such that both the fluid sources
are gravitationally coupled to each other. We have then observed
that the field equations now possess the effects of both matter
distributions that ultimately increase the degrees of freedom, and
thus, a systematic approach must be needed to solve them. We have
applied a transformation on the radial component (provided by MGD
strategy) in the field equations that split them into two distinct
systems. Each of these sets characterize their parent (perfect and
additional) sources, and we have solved them independently.

Firstly, we have dealt with the two independent field equations
associated with the perfect fluid configuration. We have observed
that the isotropic system contain four unknowns, therefore, the
Finch-Skea metric has been adopted as a perfect fluid solution to
the corresponding equations of motion. The temporal/radial
components of this metric are given by
$$e^{a_1(r)}=\frac{1}{4}\big(2C_1+C_2\sqrt{C_3}r^2\big)^2, \quad e^{a_2(r)}=C_3r^2+1,$$
comprising a triplet of unknowns $(C_1,C_2,C_3)$. We have then
considered the Schwarzschild exterior solution to match with the
interior spherical metric and calculated these constants at the
interface $\Sigma:~r=R$. On the other hand, the field equations
\eqref{g21}-\eqref{g23} representing an extra source also involve
four unknowns including a matter triplet and the deformation
function. We have tackled them by implementing three different
constraints on $\mathrm{E}$-sector separately, leading to distinct
solutions. After solving both sets of equations, we have determined
effective matter variables and anisotropy through the relations
\eqref{g13} and \eqref{g14}.

In order to ensure the physical relevancy of the developed models,
we have investigated several factors through the graphical
interpretation. We have chosen different values of the parameters
such as $\xi=0.25,0.5,0.75,1$ and $\zeta=0.1,0.2,0.3$ to check that
how the interior fluid distribution is affected by the modified
corrections and MGD scheme. The matter triplet corresponding to all
three models have been observed acceptable for every parametric
choice. We have found in each case that the lower values of $\xi$
and $\zeta$ lead to more dense structure. Furthermore, we have
plotted dominant energy conditions that ensure the presence of usual
matter in the interior of each proposed model. With reference to the
stability analysis, we have implemented the causality conditions and
cracking criterion on the resulting solutions. It is found that our
models I and II are stable for all chosen values of the model and
decoupling parameters. However, model III is stable only for
$\zeta=0.1$ and $0.2$ because the tangential sound speed criterion
has not been fulfilled for $\zeta=0.3$. It must be acknowledged here
that our results regarding models I and II are better than that
produced in the context of $\mathbb{GR}$ \cite{34}. Moreover, our
model II is compatible with the anisotropic Tolman VII \cite{25ac}
as well as Buchdahl's solutions \cite{25az}. It must be mentioned
here that putting $\xi=0$ reduces all these results to
$\mathbb{GR}$.

Finally, we have also compared our developed results with the
observational data. For this, we have presented numerical data for
each solution in Tables \textbf{1}-\textbf{9}. The energy density
has been found to be of order $10^{14}$ or $10^{15}$, which is
consistent with the requirement for the existence of compact stars.
Since we have considered a compact candidate LMC X-4, the observed
and calculated masses of this star are in agreement with each other.
The calculated values are given as
\begin{itemize}
\item $M=1.08\mathcal{M}_{\bigodot}$, $1.06\mathcal{M}_{\bigodot}$ and
$1.09\mathcal{M}_{\bigodot}$ for $\xi=0.25$ and $\zeta=0.1$
corresponding to model I, II and III, respectively.

\item $M=0.96\mathcal{M}_{\bigodot}$, $1.03\mathcal{M}_{\bigodot}$ and
$0.95\mathcal{M}_{\bigodot}$ for $\xi=0.25$ and $\zeta=0.2$
corresponding to model I, II and III, respectively.

\item $M=0.83\mathcal{M}_{\bigodot}$, $1.04\mathcal{M}_{\bigodot}$ and
$0.87\mathcal{M}_{\bigodot}$ for $\xi=0.25$ and $\zeta=0.3$
corresponding to model I, II and III, respectively.
\end{itemize}

It has been observed from the above discussion that the lower values
of both model and decoupling parameters produce a best fit to the
existing data.
\\\\
\textbf{Data Availability Statement:} This manuscript has no
associated data.

\end{document}